\documentclass[aps,prd,10pt,twocolumn,superscriptaddress,floatfix,nofootinbib,amsmath,amssymb,altaffilletter,preprintnumbers,longbibliography,eprint
]{revtex4-2}
\pdfoutput=1

\usepackage{graphicx}
\usepackage{dcolumn}
\usepackage{bm}
\usepackage{braket}
\usepackage{dsfont, fontawesome}
\usepackage[colorlinks=true, pdfstartview=FitV,  breaklinks=true]{hyperref}
\usepackage{color}
\usepackage[dvipsnames,table]{xcolor}
\hypersetup{urlcolor=BlueViolet, citecolor=Plum, linkcolor=PineGreen}
\usepackage{cleveref}
\usepackage{comment}
\usepackage{booktabs}
\usepackage{siunitx}
\usepackage{soul}
\usepackage{bm}
\usepackage{esint}
\usepackage{lettrine}
\usepackage{Zallman}

\makeatletter
\renewcommand{\@seccntformat}[1]{%
  \ifcsname prefix@#1\endcsname
    \csname prefix@#1\endcsname
  \else
    \csname the#1\endcsname\quad
  \fi}

\makeatother

\newcommand{\KCL}{\affiliation{Theoretical Particle Physics and Cosmology Group, Department of Physics, King's College London, UK}}

\begin{document}

\preprint{ \includegraphics[width=0.4\textwidth]{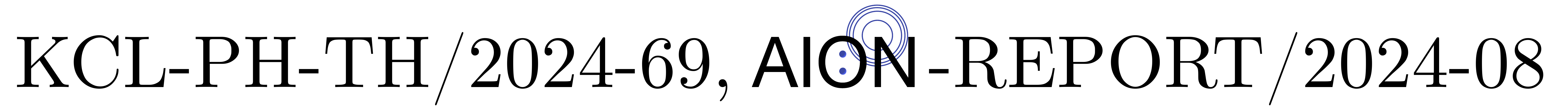}  }

\title{Massive graviton dark matter searches with long-baseline atom interferometers
}

\author{Diego Blas} 
\affiliation{Institut de F\'isica d’Altes Energies (IFAE), The Barcelona Institute of Science and Technology, Campus UAB, 08193 Bellaterra (Barcelona), Spain}
\affiliation{Instituci\'o Catalana de Recerca i Estudis Avan\c cats (ICREA), Passeig Llu\'is Companys 23, 08010 Barcelona, Spain}

\author{John Carlton}
\email{john.carlton@kcl.ac.uk}
\KCL

 \author{Christopher McCabe}
 \KCL


\begin{abstract}
Atom interferometers offer exceptional sensitivity to ultra-light dark matter (ULDM) by precisely measuring effects on atomic systems. 
Previous studies have demonstrated their capability to detect scalar and vector ULDM candidates, yet their potential for probing spin-2 ULDM remains unexplored. 
In this work, we address this gap by investigating the sensitivity of atom interferometers to spin-2 ULDM across several frameworks for massive gravity, including the Lorentz-invariant Fierz-Pauli case and two distinct Lorentz-violating scenarios.
We show that coherent oscillations of the spin-2 ULDM field induce measurable phase shifts in atom interferometers through three coupling mechanisms: scalar interactions that modify atomic energy levels, and vector and tensor effects that alter the propagation of both atoms and light.
We demonstrate that these multifaceted interactions enable atom interferometers to probe a range of ULDM properties and mass scales that are inaccessible to laser interferometric gravitational wave detectors.
Our results establish the potential of atom interferometers to open a new experimental frontier for spin-2 dark matter detection.
\end{abstract}


\maketitle

\flushbottom

\section{Introduction}

\lettrine{A}{tom} interferometry is a powerful technique for probing fundamental physics. Employing the exquisite sensitivity of cold atoms acting as quantum sensors, upcoming experiments aim to measure a range of new physics beyond the Standard Model~\cite{Buchmueller:2022djy}. Experiments have successfully measured and been proposed to measure fundamental constants~\cite{Morel:2020dww, Rosi:2014kva, Parker_2018}, test quantum mechanics~\cite{Arndt_2014, Bassi:2012bg, Manning:2015cta, Vowe:2022mzm} and general relativity~\cite{Roura:2018cfg, Zych:2011hu, Xu:2019vlt, Asenbaum_2020} to high precision, constrain fifth-forces~\cite{Biedermann_2015, Rosi:2017ieh} and models of dark energy~\cite{Hamilton:2015zga, Elder:2016yxm, Burrage:2014oza, Sabulsky:2018jma}. Several upcoming atom interferometer experiments aim to build on these measurements~\cite{Schelfhout:2024bvz, Roura:2024rmm}, and have been proposed as sensors to search for gravitational waves~\cite{Dimopoulos:2007cj, Dimopoulos:2008sv, Yu:2010ss, Chaibi:2016dze, Graham:2017pmn, Loriani:2018qej, Schubert:2019ycf, Badurina:2024rpp} and dark matter~\cite{Graham:2015ifn, Geraci:2016fva, Arvanitaki:2016fyj, Badurina:2021lwr, Badurina:2021rgt, Badurina:2022ngn, DiPumpo:2022muv}. 

Light-pulse atom interferometry is achieved by trapping and cooling atoms before splitting them into a superposition of states using precisely timed laser pulses. Atom wavepackets travel along separate paths, directed by further laser pulses, before coherently combining again at the end of the sequence. The interference patterns between the atom paths are measured, giving a precise measure of various phenomena acting on the atoms. For a more detailed review of the physical principles, see, e.g.,~\cite {Canuel:2006zz, Hu_2019, Buchmueller:2023nll}.
Several atom interferometer experiments are currently in development or have been proposed to search for new physics, including AION~\cite{Badurina:2019hst}, ELGAR~\cite{Canuel:2020cxb}, MAGIS-100~\cite{MAGIS-100:2021etm}, MIGA~\cite{Canuel:2017rrp}, and ZAIGA~\cite{Zaiga}. More ambitious space-bound projects, such as AEDGE~\cite{AEDGE:2019nxb} and STE-QUEST~\cite{Aguilera:2013uua} have also been proposed.

Among the possible areas where atom interferometry may contribute, we will focus on \emph{dark matter}. The nature of dark matter has been a long-standing mystery in cosmology and particle physics~\cite{Bertone:2016nfn}. Multiple lines of evidence, from galactic rotation curves to cosmic microwave background measurements, point to its existence~\cite{bertone2010particle}. Significant experimental efforts to directly detect dark matter have explored various mass ranges and coupling strengths. Traditional direct detection experiments searching for weakly interacting massive particles have thoroughly probed masses above $\sim\text{MeV}$, while experiments targeting axion-like particles have explored specific lighter mass windows. Despite these efforts, dark matter continues to elude direct detection, and a large portion of the possible parameter space, ranging from masses as low as $\sim 10^{-22}$\,eV~\cite{Bar:2018acw, Ferreira:2020fam, Zimmermann:2024xvd}, remains unexplored. Atom interferometer experiments have the potential to access a significant part of this unexplored parameter space~\cite{Badurina:2021rgt,TVLBAI:2024}

When considering direct searches for dark matter, besides the particle's mass, the fundamental {\emph{spin}} is also important.
While most work has focused on fundamental spin lower or equal to $3/2$, we devote this paper to exploring \emph{massive spin-2} dark matter models. 
Massive spin-2 theories have an interesting, and still unfinished, history. After the first developments for a consistent (free) Lorentz invariant theory~\cite{Fierz:1939ix}, it was soon realized that interacting massive spin-2 states had potential problems that could jeopardize any effective theory description at relatively large distances. These difficulties were partially addressed either by considering a concrete non-linear theory (dRGT)~\cite{deRham:2014zqa} or by breaking more fundamental principles, such as violating Lorentz invariance~\cite{Rubakov:2004eb,Rubakov:2008nh}. 

The first possibility has been extensively studied, and there is still controversy about its relevance as an effective field theory with a gap between massive spin-2 fields and other states, see e.g.~\cite{deRham:2014zqa,Bellazzini:2023nqj}. If the massive spin-2 candidate is not the field responsible for ordinary gravitation, but rather only constitutes the dark matter, this problem may be alleviated.\footnote{For instance, there may be scenarios where the spin-2 polarization is accompanied by more states, or where the cut-off of the theory is small, but still in agreement with DM phenomenology.} We take this agnostic attitude for this model in our work.
Furthermore, if a Lorentz-invariant massive graviton were responsible for ordinary gravitational phenomena, its mass would be constrained to be below  $\sim 10^{-23}$\,eV~\cite{deRham:2016nuf, LIGOScientific:2021sio, Poddar:2021yjd}, which is lower than the minimal mass required for particles constituting \textit{all} of the dark matter ($m\gtrsim 10^{-22}$\,eV)~\cite{Bar:2018acw, Ferreira:2020fam,Zimmermann:2024xvd}.
As a consequence, multi-metric theories (theories with more than one spin-2 excitation, see e.g.~\cite{deRham:2014zqa,Lust:2021jps}) are the only viable option for Lorentz-invariant massive spin-2 fields to constitute all dark matter.

Turning to Lorentz-violating (LV) massive gravity, this approach was spurred by the problems around the Lorentz-invariant case~\cite{Rubakov:2004eb}, and its natural emergence in the physics of gravitation in the presence of scalars \emph{spontaneously} breaking Lorentz invariance. The works~\cite{Dubovsky:2004sg,Rubakov:2008nh} showed that Lorentz-violating massive gravity is a natural effective description at low energies, which can be free from the aforementioned problems. 
Previous work~\cite{Dubovsky:2004ud,Dubovsky:2005dw} proposed a particular phase of these models containing dark matter candidates, while other options are also possible, as we will discuss. A remarkable aspect of LV massive gravity is that it can emerge from a symmetry-breaking scheme~\cite{Blas:2014ira}. Indeed, a UV completion can be found in a way similar to that for massive vectors in the Standard Model, in contrast to the Lorentz-invariant case.\footnote{This approach requires embedding the model in theories of gravitation with a preferred frame, which have been proposed as complete theories of quantum gravity~\cite{Blas:2009qj,Horava:2009uw}.} The challenge of this approach is to explain why the breaking 
of Lorentz invariance in the gravity or dark matter sector does not percolate to the standard model sector, 
where this symmetry is satisfied to very high accuracy. Some ideas on how to address this difficulty can be found in \cite{Liberati:2013xla}.
 
In this work, we focus on ultra-light dark matter (ULDM), with masses $m_{\rm DM}\lesssim 1$\,eV. 
At these mass scales, the de Broglie wavelength of the dark matter particles in the Milky Way exceeds their separation distance, and a classical field configuration, satisfying classical equations of motion, emerges~\cite{Hui:2021tkt}. This `wave-like' behaviour brings in new phenomenological opportunities to search for dark matter. For the Lorentz-invariant ULDM spin-2 case, this has been exploited in studies of its impact on gravitational wave detectors in~\cite{Armaleo:2020efr,Manita:2023mnc}, while other bounds may come from superradiant phenomena around black holes~\cite{Dias:2023ynv}. 

Building on related work for spin-0 ULDM~\cite{Badurina:2019hst,MAGIS-100:2021etm}, we investigate the sensitivity of \emph{atom interferometers} to ultra-light spin-2 dark matter candidates. We examine three viable theoretical frameworks for massive gravity -- the Fierz-Pauli Lorentz-invariant case and two Lorentz-violating scenarios -- analyzing how each couples to Standard Model fields in the non-relativistic limit relevant for atom interferometry.
We then derive sensitivity projections for upcoming long-baseline atom interferometer experiments. These proposed projects will not only probe a complementary mass range that falls between the masses probed by LISA and LIGO, but are also sensitive to additional detection channels through three mechanisms: the coupling of the scalar mode of the spin-2 field to atomic energy levels, and vector and tensor effects that alter the propagation of atoms and light.

This paper is structured as follows: Sec.~\ref{sec:mgravs} reviews massive gravity field theories as the basis of examining spin-2 dark matter. Sec.~\ref{sec:AIs} derives the signal of spin-2 dark matter in atom interferometers. Sec.~\ref{sec:limits} presents the projected sensitivity of long-baseline atom interferometer experiments to spin-2 dark matter, and details other relevant constraints. Sec.~\ref{sec:discuss} discusses the results and future research directions. 
Two appendices provide additional technical details:
App.~\ref{app:dofs} details the calculation to integrate non-propagating degrees of freedom in our theories, while App.~\ref{app:directions} examines the directional dependence of the signal.

Throughout this paper, we use the mostly $+$ metric convention $(-+++)$  and, unless explicitly included, assume $\hbar = c = 1$.

\section{Massive gravitons}\label{sec:mgravs}

Several effective field theories include massive spin-2 states at low energies~\cite{deRham:2014zqa}. 
These theories can be categorised into those generating Fierz-Pauli (FP) Lorentz-invariant mass terms at the linear level and those generating Lorentz-violating (LV) mass terms. 
While Lorentz invariance ensures that all propagating helicity modes of the new field have the same mass, this does not necessarily hold for Lorentz-violating theories, as we will demonstrate. Furthermore, the two cases allow for distinct couplings to Standard Model fields.
The following section explores these differences through the constructions of Lagrangians for massive spin-2 fields that could provide suitable dark matter candidates.

\subsection{Lorentz invariant massive gravity}

The Lorentz invariant Lagrangian for a massive spin-2 theory free of pathologies at the linearised level is the Fierz-Pauli ({\bf FP}) Lagrangian~\cite{Fierz:1939ix,deRham:2014zqa}, 
\begin{equation}
    \mathcal{L}_\mathrm{FP} = -\varphi^{\mu\nu}\varepsilon^{\rho\sigma}_{\mu\nu}\varphi_{\rho\sigma}-\frac{1}{2}m^2\left(\varphi_{\mu\nu}\varphi^{\mu\nu}-\varphi^2\right), \label{eq:FP}
\end{equation}
where $\varphi_{\mu\nu}$ is a massive spin-2 field of mass $m$, $\varphi = \varphi_\mu^{\;\;\mu}$ is the trace of the field and $\varepsilon^{\rho\sigma}_{\mu\nu}$ is the Lichnerowicz operator, defined by
\begin{equation}
\begin{split}
    \varepsilon^{\rho\sigma}_{\mu\nu}\varphi_{\rho\sigma} = &-\frac{1}{2}\big(\Box\varphi_{\mu\nu} - \partial_\mu\partial^\alpha\varphi_{\alpha\nu} - \partial_{\nu}\partial^\alpha\varphi_{\alpha\mu}\\
    &+\partial_\mu\partial_\nu\varphi-\eta_{\mu\nu}\Box\varphi+\eta_{\mu\nu}\partial_\alpha\partial_\beta\varphi^{\alpha\beta}\big),
\end{split}
\end{equation}
where $\Box := \eta^{\mu\nu}\partial_\mu\partial_\nu$. 

We work at linear level in this work and point the interested reader to \cite{deRham:2014zqa,Rubakov:2008nh} for possible non-linear completions, and the subtleties that arise with them (in particular related to the relative coefficient of the two terms in the mass term that characterizes the FP Lagrangian). As long as we consider the massive spin-2 field and the standard graviton to be independent fields, these subtleties can always be addressed, 
at least in principle.\footnote{\label{foot:MP}One of the main problems of massive gravity is the existence of a low energy cut-off $\Lambda\approx (m^x \tilde M_{\rm Pl})^{1/(x+1)}$ where $x\, \in \mathbb{N}$ and the scale $\tilde M_{\rm Pl}$ corresponds to the coupling controlling the interaction of $\varphi_{\mu\nu}$ \cite{Arkani-Hamed:2002bjr,Blas:2014aca}. Since this combination of $m$ and $\tilde M_{\rm Pl}$ is unconstrained, we will assume that it is never a problem. This problem is alleviated in the Lorentz-violating cases we discuss \cite{Dubovsky:2004sg,Rubakov:2008nh,Blas:2014ira}.} 
Furthermore, the choice of an arbitrarily small mass for the graviton is technically natural~\cite{deRham:2012ew} and therefore represents an appealing candidate for ULDM.

The coupling of the spin-2 field to matter fields will be of the form
\begin{equation}
\label{eq:coupling}
    \mathcal{L}_\mathrm{FP-SM}=\kappa^\phi \varphi^{\mu\nu}{\cal O}_{\mu\nu},
\end{equation}
where  ${\cal O}_{\mu\nu}$ is a symmetric tensor built with Standard Model fields. From the breaking of diffeomorphisms from the mass term, Eq.~\eqref{eq:FP}, there is no need for this tensor to be conserved, and tensors beyond the energy-momentum tensor can be considered. In particular,  the coupling to different species contributing to  ${\cal O}_{\mu\nu}$  is not universal, which induces violations of the equivalence principle.  This is, in spirit, similar to the violations from scalar ULDM fields, whose effect in atomic interferometers is discussed in~\cite{Geraci:2016fva, Arvanitaki:2016fyj, Badurina:2021lwr, Badurina:2021rgt, Badurina:2022ngn, DiPumpo:2022muv}. 
We postpone a more comprehensive discussion of the coupling to Sec.~\ref{sec:coupling}.

The invariance under linearized diffeomorphisms can be  restored  by including four Stückelberg fields $\chi_\mu$, as part of the
redefinition of the primordial field~\cite{deRham:2014zqa,Rubakov:2008nh}.
However, we will not make explicit use of this formalism in this work. Instead, to
 understand the dynamical degrees of freedom, we will express  $\varphi_{\mu\nu}$ as irreducible $SO(3)$ representations~\cite{Rubakov:2004eb}:
\begin{equation}
    \begin{split}
        \varphi_{00} &=  \Psi,\\
        \varphi_{0i} &=  u_i + \partial_i w,\\
        \varphi_{ij} &=  \varphi^{\rm TT}_{ij} + 2\partial_{(i}A_{j)} + \partial_i\partial_j\sigma + \delta_{ij}\pi.
        \label{eq:splitdofs}
    \end{split}
\end{equation}
Here, $\varphi^{\rm TT}_{ij}$ are the transverse-traceless tensor (under $SO(3)$) modes of the field, $u_i$ and $A_i$ are transverse vectors, $w, \sigma, \pi$ are three dimensional scalars, and 
\begin{equation}
    a_{(\mu}b_{\nu)} := \frac{1}{2}(a_\mu b_\nu+a_\nu b_\mu).
    \end{equation}
    Plugging this decomposition of the field into the FP Lagrangian in Eq.~\eqref{eq:FP}, and after integrating out the non-propagating fields (cf.\ App.~\ref{app:dofs}), 
we arrive at separate Lagrangians for the propagating degrees of freedom in each sector~\cite{Rubakov:2004eb}. Namely,
\begin{equation}
    \mathcal{L}_\mathrm{FP} = \mathcal{L}_{t}+\mathcal{L}_{v}+\mathcal{L}_{s}\,, \label{eq:allL}
\end{equation}
where the individual terms are given~by
\begin{equation}
    \begin{split}\label{eq:DOFs}
        \mathcal{L}_{t} &= \frac{1}{2}\left({\varphi}^{\rm TT}_{ij}\Box_t{\varphi}^{\rm TT}_{ij} - m_t^2{\varphi}^{\rm TT}_{ij}{\varphi}^{\rm TT}_{ij}\right),\\
        \mathcal{L}_{v}&= \frac{1}{2}\left(\widetilde{A}_{i}\Box_v\widetilde{A}_{i} -m_v^2\widetilde{A}_{i}\widetilde{A}_{i}\right),\\
        \mathcal{L}_s &= \frac{1}{2}\left(\widetilde{\pi}\Box_s\widetilde{\pi} - m_s^2\widetilde{\pi}^2\right),
    \end{split}
\end{equation}
and we have defined 
\begin{equation}
  A_i=\frac{1}{m} \sqrt{\frac{\Delta-m^2}{2 \Delta}} \widetilde{A}_i\,,~~~ \mathrm{and}  ~~~\pi = \widetilde{\pi}/\sqrt{3}.
  \label{eq:A_Pi}
\end{equation}
 The differential operators read
    \begin{equation}\label{eq:dalamb}
        \Box_\chi=-\partial_0^2+c_\chi^2 \Delta,
    \end{equation}
where $\Delta$ is the Laplace operator, $\chi=t,\,v\,\text{or}\,s$, and for the FP case $c_t=c_v=c_s=1$, and $m=m_t=m_v=m_s$. The reason for introducing a notation that allows for different masses and speeds of propagation will become clear as we turn to discuss LV massive gravity.

\subsection{Lorentz-violating massive gravity}\label{sec:LV}

Among the different ways to break Lorentz invariance, we will only discuss those that preserve the invariance under $SO(3)$ and translation of all space-time coordinates. Following Ref.~\cite{Dubovsky:2004sg}, we consider a UV scale~$F$ that characterizes the breaking of diffeomorphisms to a smaller subgroup. The kinetic term in Eq.~\eqref{eq:FP} is then completed, at leading order, by an operator term~\cite{Rubakov:2004eb,Dubovsky:2004sg,Rubakov:2008nh,deRham:2014zqa}:
\begin{equation}
\begin{split}\label{eq:LVLag}
    \mathcal{L}^\mathrm{LV}_\mathrm{mass} = \frac{1}{2}\bigg(m_0^2 \varphi_{00}^2&+2m_1^2\varphi_{0i}^2-m_2^2\varphi_{ij}^2\\
    &+m_3^2\varphi_i^i\varphi_j^j-2m_4^2\varphi_{00}\varphi_i^i\bigg),
\end{split}
\end{equation}
with $m_i\approx F^2/\tilde M_\mathrm{Pl}$, where $\tilde M_\mathrm{Pl}$ controls the interaction of $\varphi_{\mu\nu}$ (recall footnote \ref{foot:MP}). Note that this formalism allows for a well-controlled small value for the mass of the graviton~\cite{Dubovsky:2005xd}. 
The LV Lagrangian receives corrections suppressed by ${\cal O}({\rm max}[E/F,F/\tilde M_{\rm Pl}])$, where $E$ is the energy of the ULDM field. The
 latter correspond to higher-dimensional operators, such as possible modifications of the kinetic term. They are sub-leading in all cases, except when $m_0=m_1=0$. We will treat this case independently.

It is straightforward to show that when $m_1^2 = m_2^2 = m_3^2 = m_4^2=m^2$ and $m^2_0=0$, we recover the mass term in Eq.~\eqref{eq:FP}. 
The general LV case propagates up to 6 degrees of freedom, including a \emph{ghost} that makes it ill-defined. 
Several choices of parameters, protected by residual symmetries of the (linearized) diffeomorphism invariance enjoyed by the massless case, can avoid this problem. 
The residual symmetries of Eq.~\eqref{eq:LVLag} can be identified by considering the diffeomorphisms that are left unbroken for the different parameter choices. In a sense, the different choices are more robust than the FP one, as the latter is not protected by any symmetry, and its preservation by \textit{all} quantum corrections is unclear. 

We focus on \emph{phases of massive gravity} that are free from clear pathologies and protected by symmetries~\cite{Dubovsky:2004sg,Rubakov:2008nh,Comelli:2013txa,Blas:2014ira,deRham:2014zqa}. The relevant choices are:
\begin{itemize}
    \item[{\bf LV1}:] $m_1=0$ with $m_0\neq 0$. 
     This condition is protected by the residual symmetry
    \begin{equation}
        x^i\mapsto x^i+\xi^i(t). \label{eq:m10}
    \end{equation}
   Only the spin-2 polarizations propagate in this case, and they have mass $m_2$.  The expressions for the non-propagating fields are given in App.~\ref{app:dofs}. The Newtonian limit of this theory deviates from GR at large distances~\cite{Dubovsky:2004ud,Dubovsky:2005dw}. An extra scaling can be introduced to make the remaining terms satisfy the condition that makes both theories identical at large distances. 
    \item[{\bf LV2}:] $m_1=0$ and $m_0=0$. In this case, on top of the transformation in  Eq.~\eqref{eq:m10}, the protection of this choice requires the residual symmetry 
    \begin{equation}
        t\mapsto f(t), \label{eq:trepar}
    \end{equation}
    corresponding to a re-parametrization of time~\cite{Blas:2014ira}. This phase is particularly interesting, as it is the only example today of massive gravity (with or without Lorentz invariance) with a known UV completion in terms of Higgs-like fields \cite{Blas:2014ira}. Even more, it can be extended to potentially arbitrary large energies 
    by Ho\v rava gravity \cite{Blas:2014ira,Blas:2009qj,Horava:2009uw}. 
    
    To define this theory, one needs to consider also the first sub-leading operators in ${\cal O}({\rm max}[E/F,F/M_{\rm Pl}])$ to Eq.~\eqref{eq:LVLag}. These are characterized  by three new dimensionless constants $\alpha$, $\beta$ and $\lambda$ \cite{Blas:2009qj,Blas:2014aca}.\footnote{If 
    the spin-2 field mediates ordinary gravity, $\alpha,\beta,\lambda \ll 1$ \cite{Blas:2009qj,Blas:2014aca}. Newton's constant in these theories
    is modified by $O(\alpha,\beta,\lambda)$ corrections \cite{Yagi:2013ava}. As we will see, this possibility cannot be realized in nature if this field constitutes \textit{all} the dark matter.} These constants modify the properties of the propagating degrees of freedom  \cite{Blas:2014ira}. Indeed, the tensor degrees of freedom have  Lagrangian ${\cal L}_t$ in Eq.~\eqref{eq:DOFs} with 
    \begin{equation}
    c^2_t=(1-\beta)^{-1} ~~ {\rm and} ~~ m_t= m_2.
    \end{equation}
    The theory also includes a propagating scalar, that at energies below the symmetry breaking scale has~\cite{Blas:2014ira}\footnote{When comparing our results to those in~\cite{Blas:2014ira}, note that their split in scalar degrees of freedom, while similar to Eq.~\eqref{eq:splitdofs}, differs from the one here.} 
\begin{equation}
   ~~~ c^2_s\approx \frac{\lambda+\beta}{\alpha},  ~ {\rm and} ~ m_s\approx m_2\sqrt{\frac{\lambda+\beta}{2}}.
    \end{equation}
  The vector modes do not propagate in this case. The expressions for all non-propagating fields appear in App.~\ref{app:dofs}. 
    \item[{\bf LV3}:] $m_1\neq 0$ and $m_0=0$. This choice also requires $m_4=0$ to find a residual symmetry protecting it. Unfortunately, this case was identified as pathological in Ref.~\cite{Dubovsky:2004sg} from high energy completion considerations. We will not consider it any further.
    \item[{\bf LV4}:] $m_2=m_3=m_4=0$. Even if the weaker relation $m_2=m_3$ guarantees a well-behaved linear theory, the symmetry protecting this choice
    \begin{equation}
        x^i \rightarrow x^i+\xi^i(x), \label{eq:GCon}
    \end{equation}
    imposes the stronger condition $m_2=m_3=m_4=0$. This means that the spin-2 states of this phase are massless.
   The vector and scalar modes are not propagating \cite{Rubakov:2004eb,Dubovsky:2004sg}. This possibility seems interesting regarding UV completions that make the scalar sector dynamic. For instance, it is argued in \cite{Dubovsky:2004sg} that this case may have a well-defined UV completion where the scalar degree of freedom resembles the ghost condensate \cite{Arkani-Hamed:2004gbh}. As the spin-2 states are massless, we will ignore this model in our work, though it may generate new models for scalar dark matter.
\end{itemize}

When considering the possible coupling of LV massive gravity to matter, the operators in Eq.~\eqref{eq:coupling} can be written~as 
\begin{equation}
\label{eq:couplingLV}
\begin{split}
    \mathcal{L}_\mathrm{LV-SM}= &\kappa_{\rm LV}^{(2)}\varphi^{ij} {\cal O}^t _{ij} +\kappa_{\rm LV}^{(1)}\varphi^{0i}   {\cal O}^v_{i}+\kappa_{\rm LV}^{(0)}\varphi^{00}{\cal O}^s.
\end{split}
\end{equation}
This Lagrangian has extra constraints when one considers the residual subgroups of the two cases of interest. Namely, 
\begin{itemize}
    \item[{\bf LV1}:] Imposing the symmetry under
    Eq.~\eqref{eq:m10},  either $\kappa_{\rm LV}^{(1)}=0$, or 
    \begin{equation}
        \dot {\cal O}^{i}_v=\partial_j {\cal O}^{ji}_v. \label{eq:cond_m1_T}
    \end{equation}
    However, as this model does not have scalar or vector propagating degrees of freedom, we only
    consider the coupling to tensor modes in this work, while more generic tests may test all terms in Eq.~\eqref{eq:couplingLV}.
    \item[{\bf LV2}:] In this case, the extra transformation  from Eq.~\eqref{eq:trepar} implies that, on top of the  condition of Eq.~\eqref{eq:cond_m1_T} (or $\kappa_{\rm LV}^{(1)}=0$), either $\kappa_{\rm LV}^{(0)}=0$, or one needs to couple to a conserved quantity:
    \begin{equation}
    \dot {\cal O}_s=\partial_j {\cal O}^{j}_s. \label{eq:conserv00}
      \end{equation}
\end{itemize}

\subsection{Non-relativistic coupling to the Standard Model}\label{sec:coupling}
We now consider a model-agnostic approach to find the leading contributions in Eq.~\eqref{eq:coupling} and Eq.~\eqref{eq:couplingLV} in the regime relevant for atom interferometers, where non-relativistic matter and light beams are the relevant
degrees of freedom.
The couplings are suppressed by a scale $\Lambda$ characterizing the interactions of the massive gravity sector. This scale can be naturally associated with the scale $F$ introduced in Sec.~\ref{sec:LV}. 
Our discussion generalizes the possible couplings described in, e.g.,~\cite{Marzola:2017lbt,Armaleo:2020efr, Manita:2022tkl}.\footnote{See~\cite{Arkani-Hamed:2004gbh} for a discussion of couplings of the \emph{ghost condensate}.} 
We will consider the coupling directly to the atomic degrees of freedom. A more detailed analysis may distinguish the coupling to gluons or quarks, see e.g.~\cite{Arvanitaki:2014faa}. Since this paper aims to derive the first phenomenological consequences, we prefer to maintain simplicity in our analysis.

We denote the mass of the non-relativistic atomic degree of freedom (it can refer to the electrons, protons, etc.)  by $m_A$, its position by $\vec x_A$, and its three-momentum by $\vec p_A = m_A\vec v_A$. Similarly, the dark matter's speed and three-momentum are denoted by $\vec{v}_{\rm{DM}}$ and $\vec{p}_{\rm{DM}}$, respectively. In the non-relativistic limit, time derivatives yield factors of the mass, while spatial derivatives yield factors of the three-momentum.

When describing the operators for the atomic degrees of freedom, we will write them as
\begin{equation}
    {\cal O}={\cal O}^{\rm at}~\delta^{(3)}(\vec x-\vec x_A(t)).
\end{equation}

\subsubsection{Couplings of the {\bf FP}~model}

We split the possible contributions into two parts and we only consider leading contributions in velocities for each interaction:
\begin{itemize}
    \item{\it Coupling to matter}.\footnote{We ignore possible coupling to spin, as the atoms in our set-up are not polarized. Nor do we consider additional mediators that could generate $\vec{x}_A$-dependent terms.} We can write the possible coupling to the atomic momentum as
\begin{equation}
    \begin{split}\label{eq:couplingmatterFP}
   & \varphi^{\mu\nu}{\cal O}^{\rm at}_{\mu\nu}=\frac{{\alpha}^{(0)}_{\rm FP}}{\Lambda}\varphi^{\mu}_{\mu}\,m_A+\frac{{\alpha}^{(2)}_{\rm FP}}{\Lambda}\varphi^{\mu\nu}\frac{p_{A\,\mu} p_{A\,\nu}}{m_A} .
   \end{split}
\end{equation}
From Eq.~\eqref{eq:otherdofsLI} we find that the trace for free configurations vanishes,
\begin{equation}
    \varphi^{\mu}_\mu=-\Psi+\Delta \sigma +3\pi=0\;,
\end{equation}
while, from Eq.~\eqref{eq:uvsA} and Eq.~\eqref{eq:otherdofsLI}, 
\begin{equation}
\begin{split}
    & u_i \approx \frac{v_{\rm DM}}{m_A\sqrt{2}}\dot{\widetilde{A}_i},~~ \Psi\approx -2 v_{\rm DM}^2 \pi, ~~ w= 2\frac{\dot \pi}{m_A^2}.
\end{split}
\end{equation}
\begin{widetext}
As a result, the leading order interaction reads
\begin{equation}\label{eq:FPmasscoupling}
    \begin{split} 
   & \varphi^{\mu\nu}{\cal O}^{\rm at}_{\mu\nu}\approx \frac{{\alpha}^{(2)}_{\rm FP}}{\Lambda}m_A \Bigg[-2 v_{\rm DM}^2 \pi
   - 2{v_A^i}\left(2 v_{{\rm DM}\,i}\pi- \frac{v_{\rm DM}}{m\sqrt{2}}\dot{\widetilde{A}_i}\right)\\
   &~~~~~~~~~~~~~~~~~~~~+v_A^i v_A^j\left(\left(\delta_{ij}  -3 \hat v_{{\rm DM}\, i} \hat v_{{\rm DM}\,j}\right)\pi+\frac{\sqrt{2}\hat{v}_{\rm DM\, (\it i}\dot{\tilde A}_{j)}}{m}+\varphi_{ij}^{\rm TT}\right)\Bigg],
   \end{split}
\end{equation}
where  $\hat v_{\rm DM\,\it i } = v_{\rm DM\, \it i}/|\vec{v}_{\rm DM}|$ and we have used the approximations $|\vec v_{\rm DM}|\sim 10^{-3}\ll 1$ and $|\vec v_{A}|\ll1$. For $\vec v_{A}$, it can refer to the internal velocity of the electron ($\sim \alpha_{\rm em}\sim 10^{-2}$) or to the kinetic velocity of the atom~($\sim10^{-8}$). 
\end{widetext}
 \item{\it Coupling to light}. In this case, we can write
\begin{equation}
    \begin{split}\label{eq:couplinglightFP}
   & \varphi^{\mu\nu}{\cal O}^l_{\mu\nu}=\frac{{\beta}^{(0)}_{\rm FP}}{\Lambda} \varphi^{\mu}_\mu\, F^2+\frac{{\beta}^{(2)}_{\rm FP}}{\Lambda}\varphi^{\mu \nu} F_{\mu \alpha} F^\alpha_\nu . 
   \end{split}
\end{equation}
\begin{widetext}
    From the same approximations of the previous item, together with Eq.~\eqref{eq:otherdofsLI} and for a generic EM field, we have
\begin{equation}
    \begin{split}\label{eq:couplinglightFP2}
   & \varphi^{\mu\nu}{\cal O}^l_{\mu\nu}=\frac{{\beta}^{(2)}_{\rm FP}}{\Lambda}\Bigg[2F_{0i} F^{0i} v_{\rm DM}^2 \pi - 2F^{i\sigma} F^0_\sigma\left(2 v_{{\rm DM}\,i}\pi- \frac{v_{\rm DM}}{m\sqrt{2}}\dot{\widetilde{A}_i}\right)\\
   &~~~~~~~~~~~~~~~~~~~+ F^{i\sigma} F^j_\sigma\left(\left(\delta_{ij}  -3 \hat v_{{\rm DM}\, i} \hat v_{{\rm DM}\,j}\right)\pi+\frac{\sqrt{2}\hat{v}_{\rm DM\,\it (i}\dot{\tilde A}_{j)}}{m}+\varphi_{ij}^{\rm TT}\right)\Bigg].
   \end{split}
\end{equation}
\end{widetext}

\end{itemize}
Similarly, one could consider terms  of the form (e.g.\ inside covariant derivatives) 
\begin{equation}
\label{eq:inatom}
    \varphi^{\mu\nu} p_{A\, \mu} A_{\rm em\,\nu}\delta^{(3)}(\vec x-\vec x_A(t)).
    \end{equation}
These will only contribute to the laser phase (that cancels in our set-up to first order~\cite{Yu:2010ss}), or to finite-size effects that we ignore.

\subsubsection{Couplings of the {\bf LV}~models}

In this case, only scalar and tensor contributions are present. 
As before, we split the possible contribution into two parts, and we only consider leading contributions in velocities for each interaction:
\begin{itemize}
    \item{\it Coupling to matter}. In this case, we can write the matter component of   Eq.~\eqref{eq:couplingLV} as
\begin{equation}
    \begin{split}
    \label{eq:couplingmatterLV}
&   {\cal L}^{\rm at}_{\rm LV-SM}= \frac{\alpha_{\rm LV}^{(0)}}{\Lambda}m_A(\Psi +  \lambda_{\rm at} \left(\Delta\sigma+3\pi\right))\\
   &~
   + \frac{\alpha_{\rm LV}^{(1)}}{\Lambda}  p_A^i\partial_i w + \frac{\alpha_{\rm LV}^{(2)}}{\Lambda }p_A^i v_A^j\left[\partial_i\partial_j \sigma+\delta_{ij}\pi+ \varphi^{\rm TT}_{ij}\right],
    \end{split}
\end{equation}
where $\lambda_{\rm at}$ is another dimensionless constant.
For {\bf LV1}, only the tensor part is relevant. For {\bf LV2}, also the scalar part will be relevant. Note that Eq.~\eqref{eq:cond_m1_T} and Eq.~\eqref{eq:conserv00} are satisfied in this case. From Eq.~\eqref{eq:otherdofsLV} we find the strength of the different terms. 
\begin{widetext}
     \item{\it Coupling to light}. In this case, we can write
\begin{equation}
\begin{split}\label{eq:couplinglightFP3}
   {\cal L}^l_{\rm LV-SM} =
\frac{\beta_{\rm LV}^{(0)}}{\Lambda}(&-F^2+4F_{0i}F^{0i})\Psi+\frac{\beta_{\rm LV}^{(1)}}{\Lambda} F^{i\sigma} F_{0\sigma}\partial_i w\,
 \\&+
   \frac{\beta_{\rm LV}^{(2)}}{\Lambda}\left[F^{i\sigma} F^j_\sigma(\partial_i\partial_j\sigma + \delta_{ij}\pi+\varphi^{\rm TT}_{ij})\right.
  +\left(\Delta\sigma + 3\pi\right)\left.(\lambda_{l1} F^2+\lambda_{l2}F_{0i}F^{0i})\right].
   \end{split}
\end{equation}
\end{widetext}
The {\bf LV1} case only has tensorial contributions, while {\bf LV2} will include scalar and tensorial degrees of freedom. Again, we considered that Eq.~\eqref{eq:cond_m1_T} and Eq.~\eqref{eq:conserv00} must be satisfied.
\end{itemize}

\section{Spin-2 ULDM in atom interferometers}\label{sec:AIs}

\begin{figure}[t!]
    \centering
    \includegraphics[width=\columnwidth]{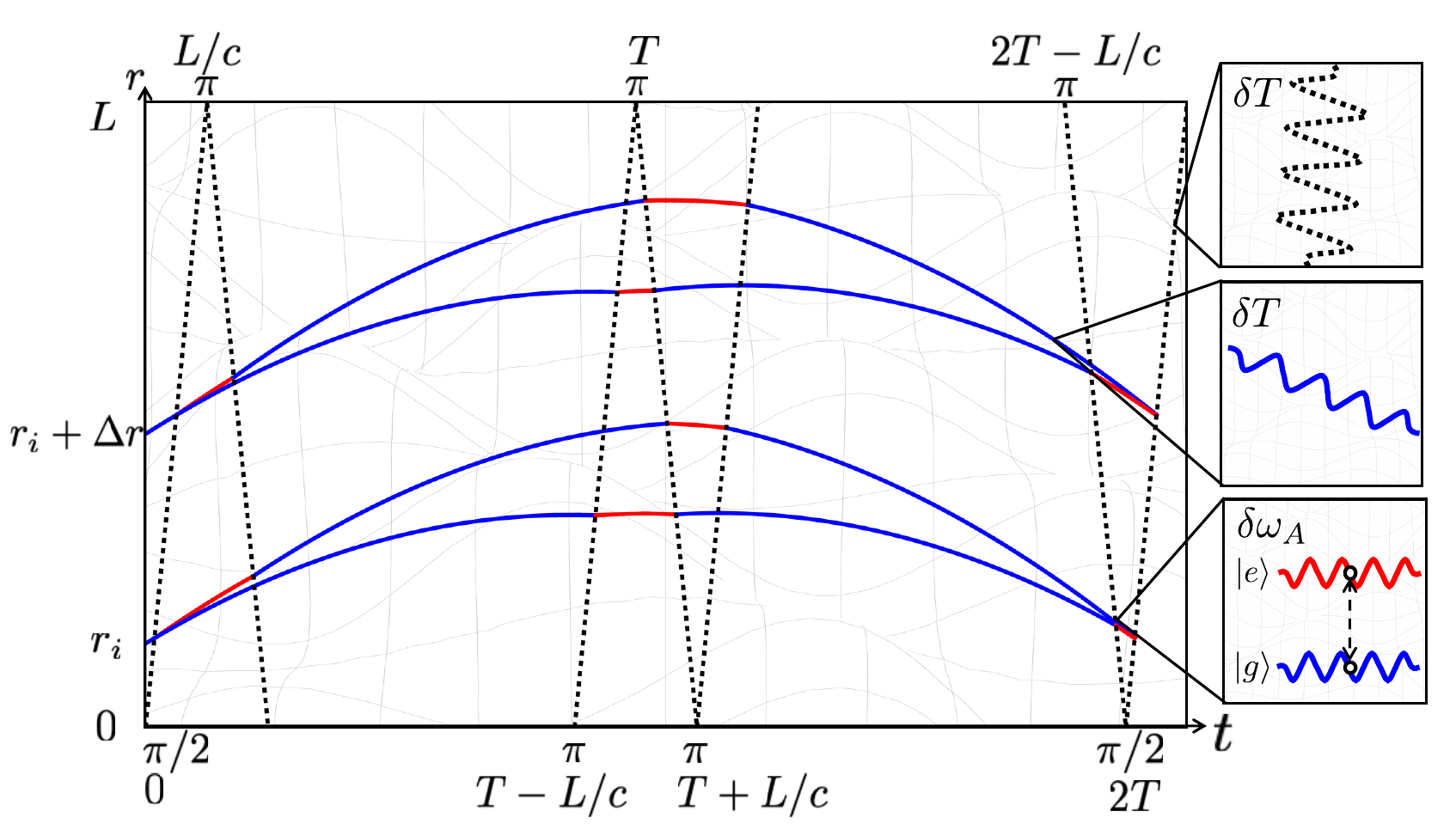}
    \caption{
    Space-time diagram of an atom gradiometer in the presence of a background spin-2 ULDM field. Two interferometers separated by $\Delta r$ share a baseline length $L$ in an LMT-enhanced Mach-Zehnder sequence (here $n=2$). The sequence consists of initial and final beam-splitters ($\pi/2$-pulses), a central mirror ($\pi$-pulse), and additional LMT $\pi$-pulses. The spin-2 ULDM field affects the interferometer through three mechanisms: tensor modes modify laser arrival times (top right), vector modes delay atom propagation (middle right), and scalar modes couple to atomic energy levels (lower right). These effects contribute to the differential phase measured between the interferometers at the sequence end.}
    \label{fig:effects}
\end{figure}

We now examine the phenomenology of the different spin-2 ULDM models in atom interferometers.
Fig.~\ref{fig:effects} shows a space-time diagram for the set-up we will consider: an atom gradiometer experiment. 
In this configuration, two (or potentially more~\cite{Badurina:2022ngn}) spatially-separated AIs are referenced by common laser sources in a fountain configuration with a large-momentum transfer (LMT) enhanced Mach-Zehnder sequence. 
During operation, two sources launch atoms that are split into a superposition of states and later redirected so they interfere at the end of the sequence.
We assume a broadband sequence for the laser pulses~\cite{Arvanitaki:2016fyj, Badurina:2021lwr, Carlton:2023ffl}. 

As we will discuss, the coherent oscillations of the background spin-2 ULDM field generates phase shifts through three mechanisms: couplings to atomic energy levels, delays in atom propagation, and delays in laser propagation. 
These effects arise from distinct interactions for each of the modes described in Eq.~\eqref{eq:coupling} or Eq.~\eqref{eq:couplingLV}. 
The expected spin-2 ULDM signal is then derived by considering how each effect impacts the many atom-laser interactions in the interferometer. 
We express our results in terms of dimensionless quantities, suppressed when necessary by~$\Lambda$.

\subsection{Dark matter configuration}\label{sec:TSDM}

The current bound on spin-2 polarizations is
$m_2\lesssim 10^{-23}\,$eV \cite{LIGOScientific:2021sio},
 while the observational limit for ULDM is $m_{\rm DM}\gtrsim 10^{-22}\,$eV~\cite{Marsh:2015xka,Ferreira:2020fam}. 
Hence, in all the cases before,  the field $\varphi_{\mu\nu}$ necessarily corresponds to a new spin-2 field (coming from a multi-gravity theory), unless it constitutes part of the total DM.  We will proceed by considering arbitrary couplings for all situations, from which one can read the constraints for the different situations.

The free action for all the cases we consider is summarized in Eq.~\eqref{eq:allL}, with the appropriate masses and speeds of propagation. We will assume that all the masses are ultra-light and that the occupation numbers are large enough to allow for a classical field description. Although we do not discuss production mechanisms for spin-2 ULDM, several options have been suggested in the past, including the misalignment mechanism~\cite{Marzola:2017lbt}, production from an anisotropic universe~\cite{Manita:2022tkl, Aoki:2016zgp, Aoki:2017cnz}, and to other possibilities discussed in~\cite{Manita:2023mnc,Dubovsky:2004ud}.\footnote{Even if not studied in that work, the model of \cite{Blas:2014ira} also allows to connect the appearance of mass to a phase transition, which may imply more
phenomenological implications.} Notice also that some of the non-propagating components of each sector may be sourced by the propagating degrees of freedom, which may be important to consider when studying the coupling to matter, cf. App.~\ref{app:dofs}.

From Eq.~\eqref{eq:DOFs}, the frequency of oscillation for each set of modes in the non-relativistic limit is  $\omega_\chi \approx m_\chi\left(1+O(|{\sigma_0}|^2)/2\right)$,  where $\sigma_0$ is the velocity dispersion of the ULDM species. We will assume $|\vec v_{\rm DM}|\approx \sigma_0\approx 10^{-3}$~\cite{Evans:2018bqy}. This sets the coherence time of 
\begin{equation}
t_{\rm coh}\sim 1/(m_\chi \sigma_0^2),
\end{equation}    for each of the modes in patches with size 
\begin{equation}
l_{\rm coh}\sim 1/(m_\chi \sigma_0).   
\end{equation}

The TT tensor modes have two polarisations labelled as $\lambda = {+,\times}$~\cite{Manita:2023mnc,Maggiore:2007ulw}, that 
generate
\begin{equation}
   {\varphi}^{\rm TT}_{i j}(t, \bm{x}) = \int {\mathrm d}^3 k \sum_\lambda \widetilde{\varphi}_{0, \lambda}(\bm{k}) e_{i j}^\lambda (\bm{k}) \cos(\omega_t t - \bm{k}\cdot\bm{x}+\phi_{t,\bm{k}}).
\end{equation}
where 
\begin{equation}
    e_{i j}^{\times}(k):=\frac{1}{\sqrt{2}}\left(p_i q_j+q_i p_j\right),    ~~ e_{i j}^{+}(k):=\frac{1}{\sqrt{2}}\left(p_i q_j-q_i p_j\right),
\end{equation}
with  $\bm{p}$ and $\bm{q}$  vectors orthogonal to each other and to~$\bm{k}$. Focusing on a coherent patch for the ULDM field for times smaller than the coherence time $t_{\rm coh}$,  and considering the non-relativistic limit, the previous equation can be rewritten in terms of two effective gradients $\bm{k}_{t,\lambda}$ and phases $\phi_{t,\lambda}$ 
\begin{equation}
    {\varphi}_{i j}^{\rm TT}(t, \bm{x}) = \sum_\lambda {\varphi}_{0, \lambda} e_{i j}^\lambda (\bm{k}_{t,\lambda}) \cos(m_t t - \bm{k}_{t,\lambda}\cdot\bm{x}+\phi_{t,\lambda}).
    \label{eq:DMfield}
\end{equation}

The $SO(3)$-vector degrees of freedom also have two polarizations. In the non-relativistic limit, we can write them as
\begin{equation}
    \widetilde{A}_{i}(t, \bm{x}) = \sum_\lambda \widetilde{A}_{0, \lambda} e_{i}^\lambda (\bm{k}_{v,\lambda}) \cos(m_v t - \bm{k}_{v,\lambda}\cdot\bm{x}+\phi_{v,\lambda}).
    \label{eq:DMfieldv}
\end{equation}
where $e_{i}^\lambda (\bm{k}_{v,\lambda})$ represent two 3-vectors orthogonal to $\bm{k}_{v,\lambda}$.  The scalar mode in the non-relativistic limit  reads
\begin{equation}
    \widetilde{\pi}(t, \bm{x}) = \widetilde{\pi}_0 \cos(m_s t - \bm{k}_{s}\cdot\bm{x}+\phi_s).
    \label{eq:DMfields}
\end{equation}

The amplitudes of the fields, and thus the signal strength, are related to the dark matter density $\rho_\mathrm{DM}$. We assume that massive gravitons (in any of the incarnations summarized by Eqs.~\eqref{eq:DOFs}) account for the entirety of the local dark matter density, which we set to the value $\rho_\mathrm{DM}=0.3~\mathrm{GeV}/\mathrm{cm}^3$~\cite{Read:2014qva}. The bounds for the case where massive gravitons only account for part of the dark matter are trivially retrieved from our results. 
We sum the contributions from the possible modes of the field to the overall dark matter density
\begin{equation}
    \rho_\mathrm{DM} = \rho_{t}+\rho_{v}+\rho_{s} = (f_{t} + f_{v}+f_{s})\rho_\mathrm{DM},
\end{equation}
where we parameterised the fractional contribution of each sector with the constants $f_\chi$. We remain agnostic about the relative values of $f_t$, $f_v$, and $f_s$, as these would depend on the specific production mechanism. For a massless sector, we assume $f_\chi=0$, since massless degrees of freedom redshift faster than massive ones.

The contributions to the energy density can be calculated from Eq.~\eqref{eq:allL} as
\begin{equation}
\begin{split}
    \rho_\mathrm{DM} &\simeq \frac{m_{t}^2}{2}\sum_\lambda  (\varphi^{\rm TT}_{0,\lambda})^{2}+ \frac{m_{v}^2}{2}\sum_\lambda (\widetilde{A}_{0,\lambda})^{2} +\frac{m_{s}^2}{2}\widetilde{\pi}_0^{2}.
\end{split}
\end{equation}
We have assumed that the polarisation tensor and vector are normalised such that $e_{i j}^\lambda e_{i j}^{\lambda\prime} = \delta^{\lambda\lambda\prime}$ and $e_i^\lambda e_i^{\lambda\prime} = \delta^{\lambda\lambda\prime}$. 
As a result,  the amplitudes of the tensor, vector, and scalar modes of the field relative to their fractional contribution to the total dark matter density are
\begin{equation}
    \begin{split}
       \varphi^{\rm TT}_{0,\lambda} = \frac{\sqrt{2f_{t}\rho_\mathrm{DM}}}{m_{t}}, \hspace{.08cm}
        \widetilde{A}_{0,\lambda} = \frac{\sqrt{2f_{v}\rho_\mathrm{DM}}}{m_{v}}, \hspace{.08cm}
        \widetilde{\pi}_0 &= \frac{\sqrt{2f_{s}\rho_\mathrm{DM}}}{m_{s}},
    \end{split}
    \label{eq:amplitudes}
\end{equation}
where when relevant, we have assumed an equal distribution in polarizations.

Having established the classical field configuration for spin-2 ULDM, we next examine how this field couples to atoms and light.
We assume all of the modes contributing to the dark matter have masses in the detection range of atom interferometer experiments, though this may not necessarily be the case.  

\subsection{Phenomenology of massive gravity in AIs}

\begin{table*}[!t]
\centering
\begin{tabular}{c|c|c|c}
\toprule
 &\text{\bf FP} & \text{\bf LV1} & \text{\bf LV2}\\
 \midrule
 ${\alpha}^{(0)}X(t)$& $-{\alpha}^{(2)}_{\rm FP}v_{\rm DM}^2\frac{\sqrt{\frac{8}{3}f_s}}{m}\cos(\phi_s(t))$ & 0 &
 ${\alpha}^{(0)}_{\rm LV}\frac{\sqrt{\frac{8}{3}f_s}}{ m_s }\left(\alpha^{-1}+\lambda_{\rm at}\right)\cos(\phi_s(t))$\\
 \hline
 $\begin{aligned}
 &\\
    &{\alpha}^{(1)}V_i(t)
 \end{aligned}$& $
     \begin{aligned}[t]
  & -2 \frac{{{\alpha}^{(2)}_{\rm FP}}}{m}\left[v_{{\rm DM},\,i}\sqrt{\frac{8f_s}{3}}\cos(\phi_s(t))\right.\\
  & ~~~~~~~~~~~~~~~~~~~~~\left.+v_{\rm DM}\sqrt{\frac{f_v}{2}}\sum_\lambda e_i^\lambda\cos(\phi_v(t))\right]
    \end{aligned}$ & 
    $\begin{aligned}
        &\\
        &0
    \end{aligned}$ & 
    $\begin{aligned}
    &\\
        &{\alpha}^{(1)}_{\rm LV}\frac{\sqrt{\frac{2}{3} f_s}}{ m_s}\frac{\hat v_{\rm DM,\it i}}{v_{\rm DM}}\frac{1+\lambda}{\lambda+\beta}\cos(\phi_s(t))
    \end{aligned}$\\
    \hline
    ${\alpha}^{(2)}M_{ij}(t)$& 0 & $\frac{{\alpha}^{(2)}_{\rm LV}}{m_t}v_A\sum_\lambda e^\lambda_{ij}\sqrt{f_t}\cos(\phi_t(t))$&0\\
\bottomrule
\end{tabular}
    \caption{Matter coupling terms in the generic Hamiltonian between the $\varphi_{\mu\nu}$ massive field and Standard Model fields (cf.\ Eq.~\eqref{eq:genericH}). We assume $\alpha^{(i)}\approx \alpha^{(j)}$ for all $i,j$. 
    For each theory, we identify at least one term that can be constrained by experimental observations. We have introduced $\phi_\chi(t)$ for the phases appearing in Eqs.~\eqref{eq:DMfield}, \eqref{eq:DMfieldv} and \eqref{eq:DMfields}.}
    \label{tab:matter}
\end{table*}

\begin{table*}[!t]
\centering
\begin{tabular}{c|c|c|c}
\toprule
 &\text{\bf FP} & \text{\bf LV1} & \text{\bf LV2}\\
 \midrule
 ${\beta}^{(0)}Y_1(t)$& 0 & 0 &  ${\beta}^{(0)}_{\rm LV}\frac{\sqrt{\frac{8}{3} f_s}}{ m_s}\left(-\alpha^{-1}+\lambda_{l1}\right)\cos(\phi_s(t))$\\
 \hline
 ${\beta}^{(0)}Y_2(t)$ & $\begin{aligned}
       \beta^{(2)}_{\rm FP}\frac{v^2_{\rm DM}}{m}\sqrt{\frac{8}{3}f_s}\cos(\phi_s(t))
   \end{aligned}$ & 0 & ${\beta}^{(0)}_{\rm LV}\frac{\sqrt{\frac{8}{3} f_s}}{ m_s}\left(4\alpha^{-1}+\lambda_{l2}\right)\cos(\phi_s(t))$\\
\hline
 ${\beta}^{(1)}W_i(t)$& 0 & 0 & $\beta_{\rm LV}^{(1)}\frac{\sqrt{\frac{2}{3} f_s}}{ m_s}  \frac{\hat v_{\rm DM,\it i}}{v_{\rm DM}}
    \frac{1+\lambda}{\lambda+\beta}\cos(\phi_s(t)) $\\
 \hline
 
 $\begin{aligned}
 \\\\
     {\beta}^{(2)}N_{ij}(t)
 \end{aligned}$ & $\begin{aligned}[t]
   \frac{{\beta}^{(2)}_{\rm FP}}{m}
    \Bigg(&\left(\delta_{ij}-3\hat v_{\mathrm{DM}\,i}\hat v_{\mathrm{DM}\,j}\right)\sqrt{\frac{2}{3}f_s}\cos(\phi_s(t))\\
    &-\hat v_{\mathrm{DM}\,i} \sum_\lambda e^\lambda_j\sqrt{2f_v} \cos(\phi_v(t))\\
   & +\sum_\lambda e^\lambda_{ij}\sqrt{f_t}\cos(\phi_t(t))\Bigg)
    \end{aligned}$ & 
    $\begin{aligned}
    \\\\
       \frac{ {\beta}^{(2)}_{\rm LV}}{m_t}\sum_\lambda e^\lambda_{ij}\sqrt{f_t}\cos(\phi_t(t))
    \end{aligned}$ & $\begin{aligned}[t]
  {\beta}^{(2)}_{\rm LV}&\left[\sum_\lambda e^\lambda_{ij}\frac{\sqrt{f_t}}{m_t}\cos(\phi_s(t))\right.\\
&\left.  +(\delta_{ij}-\hat v_i\hat v_j)\frac{\sqrt{2/3f_s}}{m_s}\cos(\phi_t(t))\right]
    \end{aligned}$\\
\bottomrule
\end{tabular}
    \caption{
    Coupling terms between the $\varphi_{\mu\nu}$ massive field and electromagnetic field in the generic Hamiltonian (cf.\ Eq.~\eqref{eq:genericH}).   
    For each theory, we identify at least one term that can be constrained by experimental observations.
    We assume $\alpha^{(i)}\approx \beta^{(i)}$ for all $i$.}
    \label{tab:light}
\end{table*}

When considering the possible experimental signatures in an atom interferometer experiment, it is more convenient to write the interactions in terms of the Hamiltonian density. We can write the generic interactions as\footnote{For Lorentz-violating operators, a term $F_{0i}\varphi^{0i}$ is \textit{a priori} possible. It is not present in our case from the symmetry of Eq.~\eqref{eq:m10}.}
\begin{widetext}
    \begin{equation}
\begin{split}
    &\mathcal{H}=\sqrt{\rho_\mathrm{DM}}\left(\frac{{\alpha}^{(0)}}{\Lambda} X(t)+\frac{{\alpha}^{(1)}}{\Lambda}  v_A^i V_{i}(t)+\frac{{\alpha}^{(2)}}{\Lambda} v_A^i v_A^j M_{ij}(t)\right)m_A\delta^{(3)}(\vec x-\vec x_A(t))\\
    &~~~~~~~~~~~+\sqrt{\rho_\mathrm{DM}}\left(\frac{{\beta}^{(0)}}{\Lambda}(F^2Y_1(t) +F_{0i}F^{0i}Y_2(t)) +\frac{{\beta}^{(1)}}{\Lambda}F_{\sigma 0} F^{i\sigma}W_i(t)+ \frac{{\beta}^{(2)}}{\Lambda} F^{i\sigma} F^j_\sigma N_{ij}(t)\right).
\end{split}\label{eq:genericH}
\end{equation}
\end{widetext}
where explicit forms for the time-dependent tensors $X(t)$, $V_i(t)$, $M_{ij}(t)$, $Y_1(t)$, $Y_2(t)$, $W_i(t)$, and $N_{ij}(t)$ are given in Tables~\ref{tab:matter} and~\ref{tab:light}. 
Different massive gravity models are characterized by different values of these tensors, which can be read off from Eq.~\eqref{eq:FPmasscoupling} and Eq.~\eqref{eq:couplinglightFP2}  for the {\bf FP} case, and from Eq.~\eqref{eq:couplingmatterLV} and Eq.~\eqref{eq:couplinglightFP3} for the {\bf LV} cases.  For the sake of conciseness, we focus on leading or particular contributions for each model where $\alpha^{(i)}\approx \beta^{(i)}$, as these may be the most relevant ones for detecting spin-2 ULDM. These (non-exclusive) operators are summarized in the aforementioned tables. The phenomenological implications of the different terms in Eq.~\eqref{eq:genericH} can be derived by adapting results from the previous literature, in particular from \cite{Graham:2012sy,Graham:2016plp, Badurina:2021lwr, Badurina:2024rpp}.

\subsubsection{\texorpdfstring{Scalar couplings $\alpha^{(0)}$ and $\beta^{(0)}$: electron mass modification and changes in energy levels}{Scalar couplings a and b: electron mass modification and changes in energy levels}}

The cases {\bf FP} and {\bf LV2} include the $\alpha^{(0)}$ term that directly modifies the mass of the Standard Model field it multiplies.  
This is similar to the leading coupling considered for scalar ULDM~\cite{Stadnik_2014}, and its most relevant effect in atom interferometers arises from the change in the electron mass:
\begin{equation}
    m_e(t) = m_{e,0}\left[1 +\frac{\alpha^{(0)}}{\Lambda}X(t)\right].
\end{equation}
As illustrated schematically in Fig.~\ref{fig:effects}, these fluctuations directly affect $\omega_A$, the transition frequency in an atom, since $\omega_A \propto m_e$~\cite{Arvanitaki:2014faa}.   There will also be a coupling to the nucleons,  but we neglect it as it is of sub-leading order, assuming all couplings are of similar strength~\cite{Arvanitaki:2014faa}. 
The signal in an atom interferometer can thus be directly inferred by adapting the results from previous work (see, e.g.,~\cite{Badurina:2021lwr}). 

Additionally, the scalar coupling $\beta^{(0)}$ directly modifies the electromagnetic coupling constant.\footnote{The coupling $\beta^{(2)}$ may have a similar impact, but in an anisotropic way. We ignore it in the following.} Note that this is the case for both $Y_1(t)$ and $Y_2(t)$, since they both modify the coupling to $\vec E^2$.
Pairing the effect from the coupling to electron mass and fine structure, and assuming $\alpha^{(0)}\approx\beta^{(0)}$ the atomic transition frequency in an atom is modified by
\begin{equation}
    \omega_A(t) \simeq \omega_{A,0}\left[1+\frac{\alpha^{(0)}}{\Lambda
    }X(t)+\frac{\beta^{(0)}}{\Lambda}\left(2+\xi_A\right)Y(t)\right],
    \label{eq:tdep_freq}
\end{equation}
where $Y(t)$ could be either $Y_1(t)$ or $Y_2(t)$, and $\xi_A\approx0.06$ for the clock transition in $^{87}$Sr that we consider~\cite{Angstmann:2004zz}.

\subsubsection{\texorpdfstring{Tensor couplings $\alpha^{(2)}$ and $\beta^{(2)}$:\\ photon propagation delay}{Tensor couplings a and b: photon propagation delay}}

The terms $\alpha^{(2)}$ and $\beta^{(2)}$ in Tables~\ref{tab:matter} and~\ref{tab:light} represent an acceleration to the trajectories of matter and light degrees of freedom, respectively.
The sensitivity to similar couplings of tensor modes has been well established in previous works~ \cite{Graham:2016plp,Badurina:2024rpp}. Assuming that $\alpha^{(2)}\approx \beta^{(2)}$, the leading contribution comes from the modification to the time of flight of the photons coming from $\beta^{(2)}$, which modifies the times at which the $\pi/2$- and $\pi$-pulses interact with the atoms, as light will move on geodesics of the effective metric $N_{ij}$. This is illustrated schematically in the upper right panel of Fig.~\ref{fig:effects}. Notice that the matrix $N_{ij}$ is still traceless, though not transverse in general.
The $\alpha^{(2)}$ term will induce a similar effect to the atom propagation delay discussed in the next subsection for the $\alpha^{(1)}$ term. However, it will be suppressed by the atom velocity since $v_A\ll1$.  

Explicitly the phase is calculated for the $k$-th path segment as
\begin{equation}
    \delta\Phi_k =  \int L_k\,\mathrm{d}t =  \int m_k\,\mathrm{d}s = m_k\delta\tau_k,
    \label{eq:phase-seg}
\end{equation} 
where $m_k$ is the mass of the atoms given by $m_A$ in the ground state, and $m_A+\omega_A$ in the excited state and $\delta\tau_k$ is the duration of the segment. Here, $\omega_A$ is the frequency of the transition used to perform the interferometry. Thus a delay in the laser propagation $\delta\tau^{(2)}$ directly leads to a measurable phase \cite{Graham:2012sy,Graham:2016plp}.

\subsubsection{\texorpdfstring{Vector couplings $\alpha^{(1)}$ and $\beta^{(1)}$:\\ atom propagation delay}{Vector couplings a and b: atom propagation delay}} \label{sec:vec_coup}

The vector couplings, corresponding to $\alpha^{(1)}$ and $\beta^{(1)}$, represent less studied effects for gravitational waves and ULDM in atom interferometer experiments. 

The term multiplying $\beta^{(1)}$ represents a coupling of the ULDM background to $\epsilon^{ijk} E_j    B_k$. This is reminiscent of the coupling of axions and is expected to generate birefringence effects~\cite{Carroll:1989vb}. These effects have been discussed as a way to detect axions detected in cavity experiments, e.g.~\cite{Liu:2018icu,Obata:2018vvr}, but they have not been studied in detail for atom interferometers and we leave them for future work. 

The term represented by  $\alpha^{(1)}$ is present in the generic study of the phase shifts in atomic interferometers for generic metrics performed in~\cite{Badurina:2024rpp}. It yields a similar effect to the tensor modes, delaying the propagation of the atoms along their trajectories and imprinting a phase when laser light interacts. It is represented schematically by the middle box in Fig.~\ref{fig:effects}. The correction to the atom's three-velocity can be derived by adapting the solution for the geodesic equations from~\cite{Badurina:2024rpp}. Indeed, from the coupling $v^ih_{i0}$ of Ref.~~\cite{Badurina:2024rpp}, we find 
\begin{equation}
\begin{split}
    \delta v^i 
    &=-\eta^{ij}\int_{t_0}^t{\rm d}t^\prime~\partial_0h_{j0},
\end{split}
\label{eq:dv}
\end{equation}
which in our case implies
\begin{equation}
    \delta v_i \simeq \frac{\alpha^{(1)}}{\Lambda} V_i(t).
\end{equation}
The delay in atom-light interaction proper time is then found by integrating this velocity over a path segment of the atom interferometer sequence
\begin{equation}
    \delta\tau^{(1)}_i=\int_{t_0}^t{\rm d}t^\prime~\delta v_i.
\end{equation}

The delay depends on which of the theories we consider. The two relevant ones are:
\begin{itemize}
    \item{\bf FP}: in this case, from Table~\ref{tab:matter}, one notices that the leading order delay is suppressed by a term $v_{\rm DM}$, suggesting it may not contribute to the signal if the couplings to matter and light are all of the same order of magnitude.

    \item{\bf LV2}: in this case,  from Table~\ref{tab:matter},  the coupling to vectors is enhanced by $v_{\rm DM}$. As a result, this can be a leading contribution and we consider it in the following.
\end{itemize}

\subsection{Signal in an atom interferometer}\label{sec:signal}

Having established both the ULDM field configuration and the possible interactions, we now derive the signal in an atom interferometer experiment. This calculation combines the three effects discussed above -- modifications to atomic energy levels, atom propagation delays, and laser propagation delays -- into a single measurable phase shift.

 A laser pulse emitted at $r_0$ and time $t_0$ intersects atoms at $r$ at a time modified by delays to either the laser propagation or atom propagation
\begin{equation}
    \tau(r,r_0,t_0) = t_0 + |r-r_0| + \delta\tau^{(1)} + \delta\tau^{(2)}.
\end{equation}
    In each case, the proper time will take the form
\begin{equation}
\begin{split}
    \tau(r, r_0, t_0) = t_0 &+ |r-r_0| \\
    &+\frac{\gamma_\chi}{m_\chi}\bigg[\sin\left(m_\chi |r-r_0|+m_\chi t_0+\phi_\chi\right) \\
    &\qquad\quad -\sin\left(m_\chi t_0+\phi_\chi\right)\bigg],
\end{split}
\end{equation}
where $\gamma_\chi$ represents the leading coupling for the $\alpha^{(1)}$ or $\beta^{(2)}$ terms in Tables~\ref{tab:matter} and~\ref{tab:light}, and where we have integrated the factor of $\cos(\phi(t))$.

The difference in the laser intersect time for two interferometers at $r=0$ and $r=L$ is given by
\begin{equation}
\begin{split}
    \Delta\tau(t) = L
    +\frac{\gamma_\chi}{m_\chi}\bigg[&\sin\left(m_\chi (t+L)+\phi_\chi\right)\\
    &~~~~~~~~~~~~~~-\sin(m_\chi t+\phi_\chi)\bigg].
\end{split}
\label{eq:proptdiff}
\end{equation}
In an interferometer sequence featuring $n$ large momentum transfer pulses per interaction, the emission times of light pulses are $t_{q,i}=t_0+iL$. Combining the time-dependent transition frequency from Eq.~\eqref{eq:tdep_freq}, and Eq.~\eqref{eq:proptdiff} into a single quantity $[\omega_A\Delta\tau](t) = \omega_A(t)\Delta\tau(t)$, the gradiometer phase is
\begin{equation}
\begin{split}
    \Delta\Phi = \sum_{i=0}^{n-1}\bigg(&[\omega_A\Delta\tau](t_{2,-i}) - [\omega_A\Delta\tau](t_{1,i})\\
    -&[\omega_A\Delta\tau](t_{1,-i})+[\omega_A\Delta\tau](t_{0,i})\bigg).
\end{split}
\end{equation}
Keeping the terms only linear in $1/\Lambda$, and in the limit $m_\chi L \ll 1$ we arrive at the differential phase shift 
\begin{widetext}
\begin{equation}
\begin{split}
     \Delta\Phi=\sum_\chi\Delta\Phi_\chi
     = &\sum_\chi4\gamma_\chi\frac{\omega_{A,0}}{\Lambda}\frac{\sqrt{\rho_{\rm DM}}}{m^2_\chi} \frac{\Delta r}{L} \sin\left[\frac{m_\chi  n L}{2}\right]   \sin\left[\frac{m_\chi  T}{2}\right]
     \sin\left[\frac{m_\chi  (T-(n-1)L)}{2} \right]\cos\left[m_\chi\frac{2T+L}{2} + \phi_\chi\right],
\end{split}
\label{eq:signal}
\end{equation}
\end{widetext}
where the difference between the three signal sources we consider enters via the mass (angular frequency) of the mode $m_\chi$, the random phase $\phi_\chi$ introduced in Eqs.~\eqref{eq:DMfield}, \eqref{eq:DMfieldv} and \eqref{eq:DMfields},  and $\gamma_\chi$ corresponding to the terms in Table~\ref{tab:gamma}.\footnote{We note this phase matches the form of equations derived in previous work~\cite{Graham:2016plp, Badurina:2021lwr}.} 

In each case we assume the leading constraint to be on the $\beta^{(2)}$ term. Except for the {\bf LV2} case where leading contributions also come from the $\alpha^{(0)}$, $\beta^{(0)}$, and $\alpha^{(1)}$ terms. In laser interferometers such as LIGO and LISA each of these terms may also be constrained. However, we generally assume the $\beta^{(2)}$ term to be leading. For these experiments the strain signal in the detector will be~\cite{Manita:2023mnc}
\begin{equation}
    h(t) = \frac{\beta^{(2)}}{\Lambda}\frac{\sqrt{f_t \rho_{\rm DM}}}{m_t}\sum_\lambda D^{ij}e_{ij}^\lambda \cos(\phi_t(t)).
\end{equation}

The angular dependence of the tensor and vector signals depend on the detector tensor $D^{ij}$ and vector $D^{i}$, which are discussed further in App.~\ref{app:directions}.

\begin{table*}[!t]
\centering
\begin{tabular}{c|c|c|c}
\toprule
 $\gamma_\chi$&\text{\bf FP} & \text{\bf LV1} & \text{\bf LV2} \\
 \midrule
$\alpha^{(0)}$ & $-\alpha^{(0)}v_{\rm DM}^2\frac{8\sqrt{2f_s}}{\sqrt{3}}$ & 0 & $\alpha^{(0)}\frac{8\sqrt{2f_s}}{\sqrt{3}}(\alpha^{-1}+\lambda_{\rm at})$ \\
\hline
$\beta^{(0)}$ & $\beta^{(0)}v^2_{\rm DM}\frac{8\sqrt{2f_s}}{\sqrt{3}}$ & 0 & $\beta^{(0)}\frac{8\sqrt{2f_s}}{\sqrt{3}}(3\alpha^{-1}+\lambda_{l1}+\lambda_{l2})(2+\xi_A)$ \\
\hline
$\alpha^{(1)}$ & $-2\alpha^{(1)}v_{\rm DM}D^i\bigg[\hat v_{\rm DM,\,\it i}\sqrt{\frac{8f_s}{3}}+\sqrt{\frac{f_v}{2}}\sum_\lambda e^\lambda_i\bigg]$ & 0 & $\alpha^{(1)}\frac{2\sqrt{f_s}}{\sqrt{3}}D^i\frac{\hat v_{\rm DM,\it i}}{v_{\rm DM}}\frac{1+\lambda}{\lambda+\beta}$ \\
\hline
$\beta^{(1)}$ & 0 & 0 & 0 \\
\hline
$\alpha^{(2)}$ & 0 & 0 & 0 \\
\hline
 $\beta^{(2)}$ & $\begin{aligned}
 &\beta^{(2)}D^{ij}\Big[(\delta_{ij}-3\hat v_{\rm DM,\it i} \hat v_{\rm DM, \it j})\sqrt{f_s}\\
           &\qquad+\sqrt{f_t}\sum_\lambda e_{ij}^\lambda-\sqrt{f_v}\hat v_{\rm DM,\it i}\sum_\lambda e_j^\lambda\Big]
           \end{aligned}$ & $\beta^{(2)}\sqrt{f_t}\sum_\lambda D^{ij}e_{ij}^\lambda$ & $\begin{aligned}&\beta^{(2)}D^{ij}\Big[\sqrt{f_t}\sum_\lambda e_{ij}^\lambda\\
         &\quad+\frac{2\sqrt{f_s}}{\sqrt{3}}(\delta_{ij}-\hat v_{\rm DM,\it i} \hat v_{\rm DM,\it j})\Big]\end{aligned}$\\
         \bottomrule
\end{tabular}
    \caption{Spin-2 coupling terms $\gamma_\chi$ (see Eq.~\eqref{eq:signal}) for each of the cases we consider. We neglect the $\alpha^{(2)}$ terms as we expect them to always be subleading with respect to the $\beta^{(2)}$ terms.}
    \label{tab:gamma}
\end{table*}

\section{Sensitivity Projections}\label{sec:limits}

We now analyse the sensitivity of atom interferometers to spin-2 ULDM signals by calculating projected detection limits with experimental parameters based on conservative estimates for future experiments, such as AION and MAGIS~\cite{Badurina:2019hst,MAGIS-100:2021etm, TVLBAI:2024}. 
These limits depend on both the signal strength, derived in Sec.~\ref{sec:AIs}, and the experimental noise characteristics. 
For each of the couplings identified earlier, we determine the minimum detectable coupling strength as a function of the ULDM mass. 
We first establish the signal-to-noise framework for our analysis, which closely follows the discussion in~\cite{Badurina:2021lwr}, and then present projected sensitivities for three example experiments with different baseline lengths.

\begin{figure}[t]
    \centering
    \includegraphics[width=0.8\columnwidth]{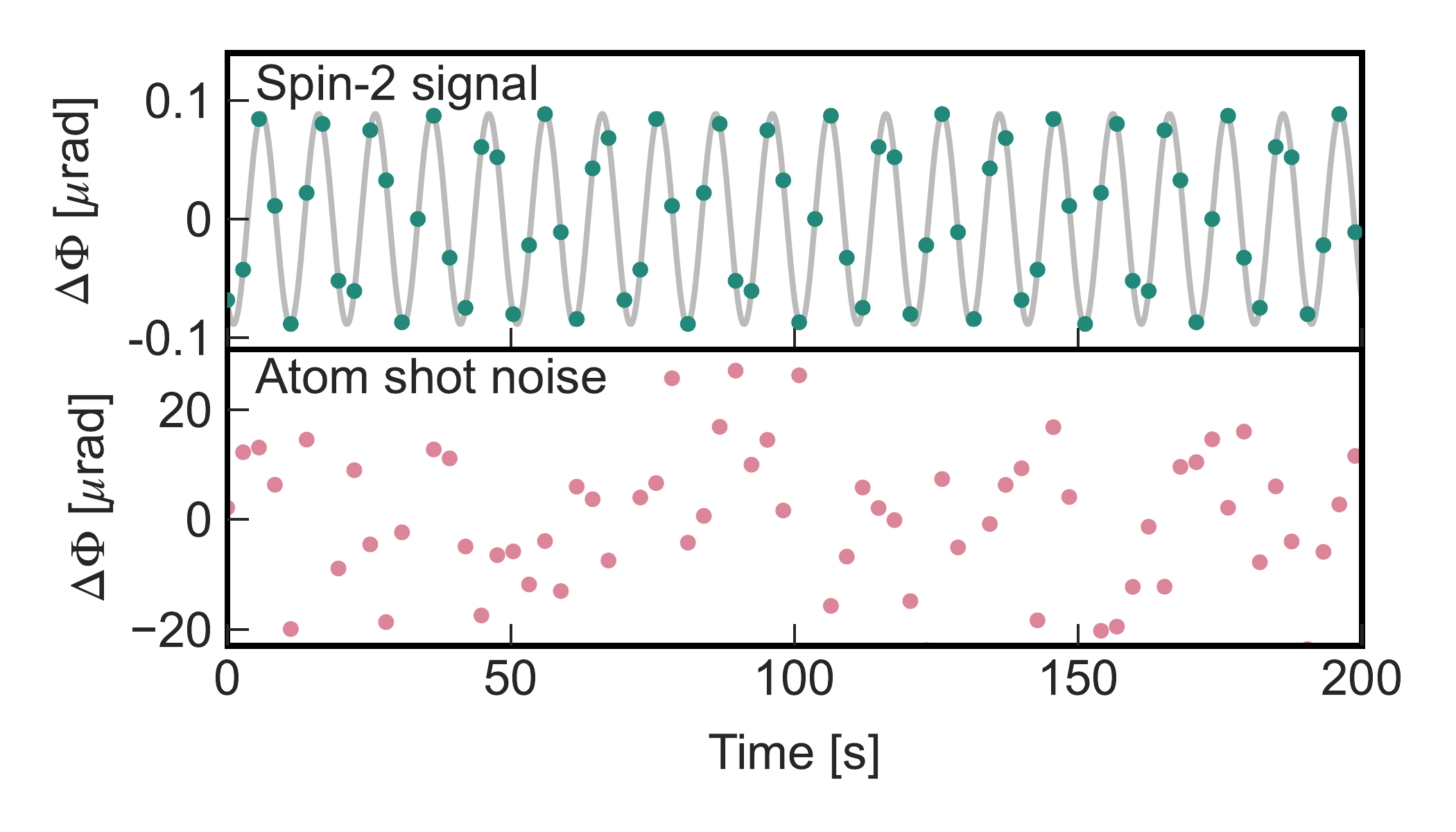}
    \includegraphics[width=0.8\columnwidth]{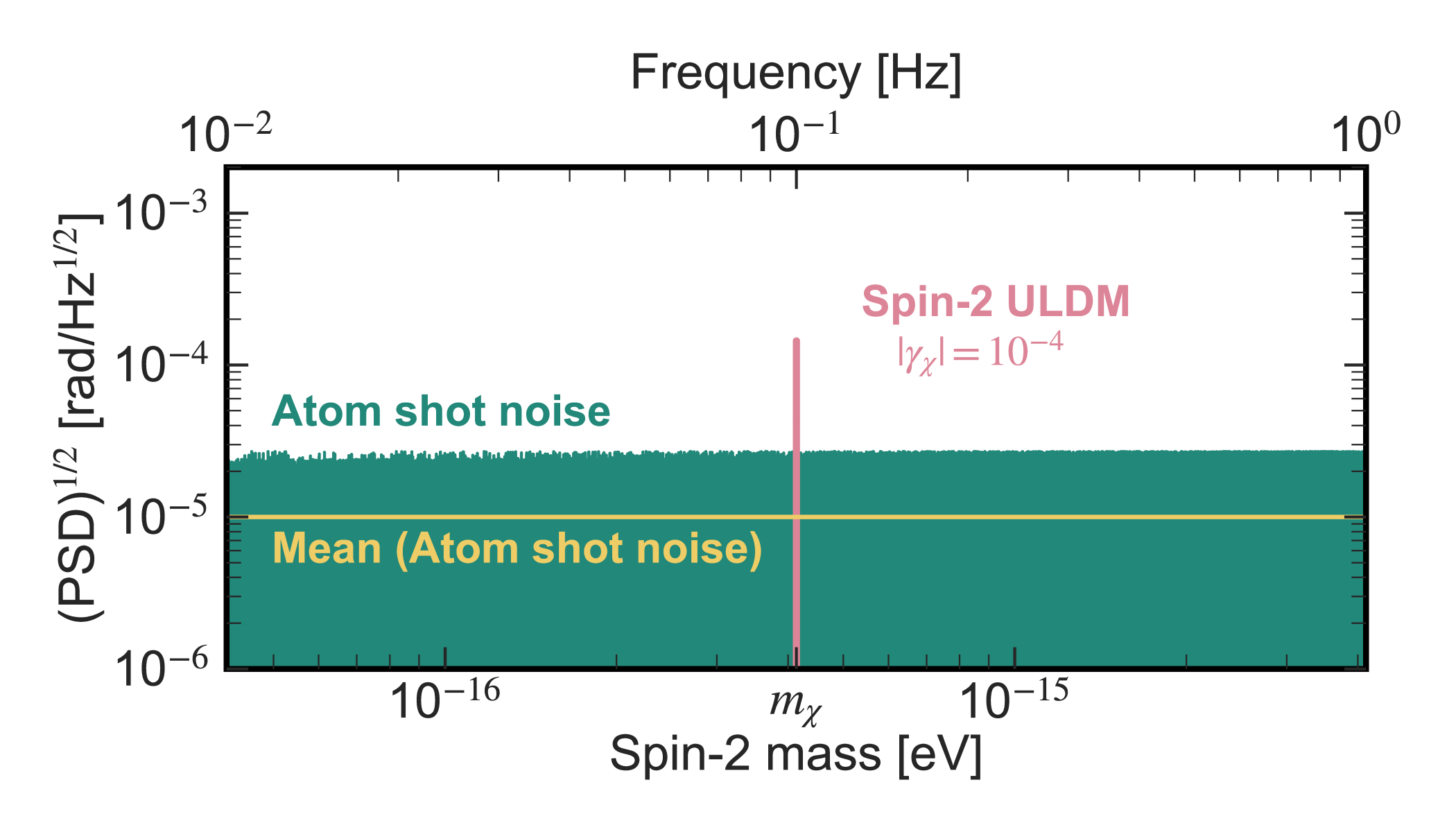}
    \caption{Upper panel: Gradiometer phase difference between the upper and lower atom interferometers from an oscillating spin-2 ULDM field (upper) and from atom shot noise (ASN, lower). Lower panel: Square-root of the PSD for the ULDM phase signal (pink spike) and atom shot noise (green curve). The yellow line shows the root-mean-square for atom shot noise, which is constant in frequency space. The parameters for the signal and noise are the same in both panels and are based on the $L=100$\,m atom interferometer parameters given in Table~\ref{tab:params}. We assume the ULDM has a mass (frequency) $\sim 4\times10^{-16}$\,eV ($0.1$\,Hz) and coupling $|\gamma_\chi| = 10^{-4}$.}
    \label{fig:DM_SN}
\end{figure}

The upper panel of Fig.~\ref{fig:DM_SN} shows an example time series of differential phase measurements $\Delta\Phi$ for an oscillating spin-2 ULDM signal (upper) and atom shot noise (lower). As the signal is much smaller than the noise, a long integration time is required to detect it. We define the square of the signal-to-noise ratio (SNR) as the power spectral density (PSD) of the signal over the noise PSD:
\begin{equation}
    \mathrm{SNR}^2 = \frac{S_s(\omega)}{S_n(\omega)}.
\end{equation}
We assume each experiment is atom-shot-noise-limited (white noise) with no additional gravity gradient noise (coloured noise). While this assumption may be difficult to achieve across the whole frequency spectrum, it is the target for the mid-band frequency range between LIGO and LISA.\footnote{For further discussion on gravity gradient noise and mitigation in these experiments, see, e.g.,~\cite{Badurina:2022ngn, Carlton:2023ffl, Carlton:2024lqy, Mitchell:2022zbp, Bertoldi:2024}.} The shot-noise PSD is flat across all frequencies and equal to the variance
\begin{equation}
    S_n = \sigma^2_{S_n} = \frac{2 \Delta t}{C^2 N_a},
\end{equation}
where $N_a$ is the number of atoms per shot in the experiment, $C$ is the contrast (we assume $C=1$), and $\Delta t$ is the repetition rate of the experiment~\cite{Badurina:2021lwr}. The values of $S_n$, together with the other experimental parameters that we consider, are listed in Table~\ref{tab:params}. The simulated square-root of the PSD is shown in the lower panel of Fig.~\ref{fig:DM_SN}, with parameters taken from the middle row of Table~\ref{tab:params}. The atom-shot-noise is flat across frequency space, with a mean corresponding to $\sigma_{S_n}$, while the ULDM signal appears as a narrow spike at a frequency set by the ULDM mass.

\begin{table}[ht]
    \centering
    \begin{tabular}{c | c | c | c | c | c | c} 
    \toprule
    Isotope  & $L$ [m] & $T$ [s] & $n$ & $\Delta r$ [m] & $S_n$ [$\mathrm{Hz}^{-1}$] & $T_\mathrm{int}$ [yrs] \\ 
    \midrule
    ${}^{87}$Sr & 10 & 0.74 & 1000 & 5 & $10^{-8}$ & 1 \\
    \hline
    ${}^{87}$Sr & 100 & 1.4 & 1000 & 90 & $10^{-10}$ & 1\\ 
    \hline
    ${}^{87}$Sr & 1000 & 1.4 & 1000 & 980 & $0.09\times 10^{-10}$ & 1 \\ 
    \bottomrule
    \end{tabular}
    \caption{Atom interferometer parameters used in our calculation of ULDM detection limits for three example long-baseline experiments. We base these parameters on conservative estimates for future experiments such as AION and MAGIS~\cite{Badurina:2019hst,MAGIS-100:2021etm, TVLBAI:2024}. The isotope of the atoms used is given; $L$ is the baseline length separating two atom sources; $T$ is the interrogation time between interferometer pulses; $n$ is the number of large-momentum transfer pulses; $\Delta r$ is the separation between the upper interferometer and noise source; $\delta\phi$ is the expected level of atom shot noise in the experiment; and $T_\mathrm{int}$ is the total run time (integration time) of the experiment.}
    \label{tab:params}
\end{table}

\begin{figure}[t!]
    \centering
    \includegraphics[width=\columnwidth]{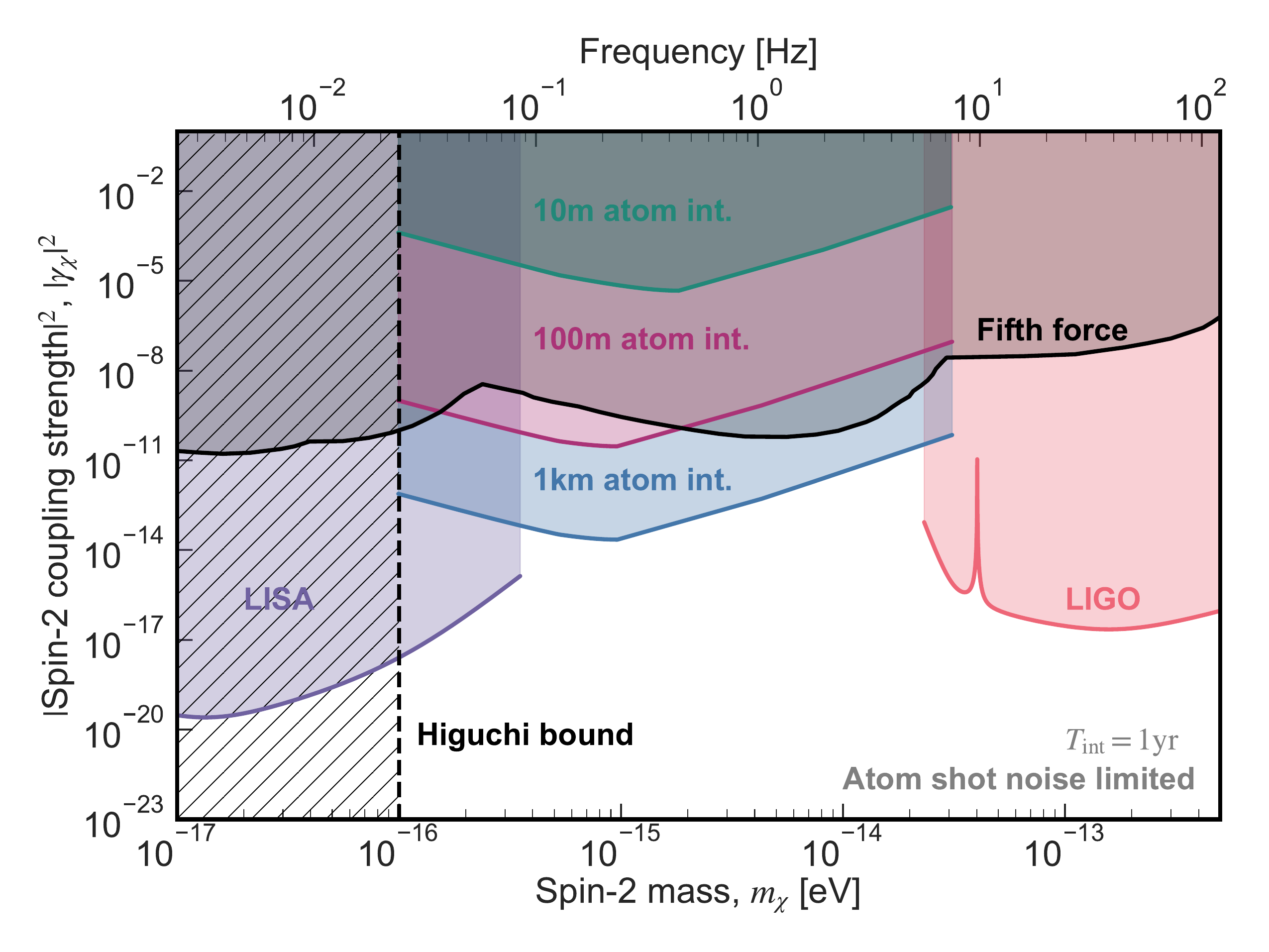}
    \caption{Sensitivity projections for detecting spin-2 ULDM signals with respect to the coupling strength $|\gamma_\chi|^2$ defined in Table~\ref{tab:gamma}. Three example atom interferometer experiments are shown of increasing baseline lengths ($\SI{10}{\meter}, \SI{100}{\meter}, \SI{1}{\kilo\meter}$), and LIGO~\cite{LIGO:2018} and LISA~\cite{Robson:2018ifk} to compare. We assume a 1-year measurement campaign for SNR~$=1$ and that each experiment is atom shot noise limited. Leading constraints come from fifth-force experiments, predominantly from lunar laser ranging~\cite{Adelberger:2003zx} and planetary measurements~\cite{Sereno:2006mw}. These are model dependent as they only constrain the scalar couplings.}
    \label{fig:FP_limits}
\end{figure}

The signal PSD is determined by averaging over the phases $\phi_{\chi,\,\lambda}$, and the spatial average of the square of Eq.~\eqref{eq:signal} (see App.~\ref{app:directions} for details). The power of the signal will scale proportional to the integration time $T_\mathrm{int}$, thus giving 
\begin{equation}
   S_s(\omega) \approx T_\mathrm{int} \langle\Delta\Phi^2_\chi\rangle.
\end{equation}

To examine the sensitivity of an experiment in terms of the couplings to Standard Model fields, we factor~$\gamma_{\chi}$ out of the signal in Eq.~\eqref{eq:signal} and define
\begin{equation}
        \langle\Delta\Phi^2_\chi\rangle = |\gamma_\chi|^2 \langle\Delta\Phi_{r}^2\rangle, 
\end{equation}
where $\gamma_\chi$ corresponds to the terms in Table~\ref{tab:gamma}.
Thus, we can express how each of the couplings scales with experimental parameters
\begin{equation}
    |\gamma_\chi|^2 \simeq \frac{\mathrm{SNR}^2}{\langle\Delta\Phi_{ r}^2\rangle}\frac{S_n}{T_\mathrm{eff}}.
\end{equation}
Here, we have replaced the integration time with $T_\mathrm{eff}$, defined as
\begin{equation}
    T_\mathrm{eff} = 
    \begin{cases}
        T_\mathrm{int} & \text{for } T_\mathrm{int} \leq \tau_c,\\
        \sqrt{T_\mathrm{int}\tau_c} & \text{for } T_\mathrm{int} > \tau_c,
    \end{cases}
\end{equation}
where $\tau_c$ is the coherence time of the ULDM field. This expression follows from an application of Bartlett’s method, which allows us to compute sensitivity projections without requiring detailed assumptions about the ULDM speed distribution (see, e.g.,~Refs.~\cite{Arvanitaki:2016fyj,Badurina:2019hst,MAGIS-100:2021etm,Badurina:2021rgt, Carlton:2023ffl}).\footnote{A likelihood analysis would be needed to extract detailed properties of the ULDM field in the event of a discovery~\cite{Badurina:2023wpk}. }

Figure~\ref{fig:FP_limits} shows the projected sensitivity of three example long-baseline atom interferometer experiments to spin-2 ULDM, using the parameters detailed in Table~\ref{tab:params}. 
For comparison, we also show the leading experimental constraints from fifth-force experiments. These constraints only apply for the scalar couplings or in the {\bf FP} case where we expect all modes to have the same mass. The tensor and vector couplings may evade these constraints if the scalar mode is much more weakly coupled or lives in a different part of parameter space. The sensitivity of LIGO and projected sensitivity of LISA are plotted to compare, and the Higuchi stability bound on the spin-2 ULDM mass, which only applies in the {\bf FP} case. 

For simplicity, in each plot we assume $\Lambda = M_{\rm Pl}$. As stated earlier, we assume the detectable parts of the field are the entirety of the dark matter, and the bound on $\gamma_\chi$ applies to each term independently.

Each of the example atom interferometer experiments uses the isotope strontium-87. Several upcoming experiments propose using this isotope for the clock transition~\cite{Badurina:2019hst, MAGIS-100:2021etm}, which gives a transition frequency $\omega_A \approx 2.7\times 10^{15}$~Hz. We expect other suitable isotopes to couple to the ULDM in a similar way, with a different numerical value for the transition frequency~\cite{Zhou:2024qdw}.

\subsection{Other constraints on spin-2 ULDM}

The scalar modes of a massive spin-2 ULDM field modifies the local Newtonian gravitational potential by a term of the form~\cite{Marzola:2017lbt} 
\begin{equation}
    \delta V_\mathrm{Newt} \propto (\alpha^{(0)})^2 e^{-m_s r},
\end{equation}
where $r$ is the separation between two objects. In the parameter space, we examine for interferometer experiments, the strongest constraints come from lunar and planetary ranging~\cite{Adelberger:2003zx, Sereno:2006mw, Armaleo:2020efr}. The scalar modes are also constrained by equivalence principle tests such as the MICROSCOPE experiment~\cite{MICROSCOPE:2022doy}, but these are not as strong as the fifth force constraints. The fifth force constraints are plotted in Fig.~\ref{fig:FP_limits}, where the scalar mode is assumed to have a mass in this parameter space. These do not necessarily constrain the vector and tensor couplings unless they live in the same part of parameter space as the scalar mode, and have a similar coupling strength. This behaviour is modified at large distances in {\bf LV2}~\cite{Blas:2014ira}.

Bosonic ULDM candidates can also be constrained by black hole superradiance. The presence of a spin-2 field around spinning black holes may trigger an instability, absorbing angular momentum and forming a cloud-like condensate that emits gravitational waves~\cite{Brito:2013wya, Brito:2020lup, Dias:2023ynv}. Future atom interferometer experiments will place superradiance constraints on ULDM in the deci-Hertz range in addition to the limits derived here from direct detection.

Another important limit for massive spin-2 is the Higuchi bound \cite{Higuchi:1986py}. This is a \emph{lower} bound 
\begin{equation}
    m^2\geq 2 H^2,
\end{equation}
of the mass of spin-2 particles with {\rm FP} mass term and propagating in a universe with a positive cosmological constant ($H^2=\Lambda_{\rm cc}/3$), where $H$ is the Hubble parameter.
For generic isotropic and homogeneous cosmological backgrounds and Lorentz-breaking situations, the equivalent bound is derived in \cite{Blas:2009my} (see also \cite{Fasiello:2013woa}). The relevance of this bound depends on the time at which the massive graviton is generated, as for masses below the (generalized) Higuchi bound one gets instabilities. The most conservative option would be to prevent all instabilities from the epoch of Big Bang Nucleosynthesis~\cite{Jain:2021pnk}, implying a minimum mass $m\sim T_\mathrm{BBN}/M_\mathrm{Pl}^2 \sim 10^{-16}$~eV, though this is a model-dependent statement. We only show the related bound for the {\bf FP} case.

\section{Discussion and outlook}\label{sec:discuss}

Long-baseline atom interferometer experiments offer exciting prospects for measuring ultra-light dark matter and gravitational waves in the frequency mid-band between LIGO and LISA. This work has identified a new theory target for long-baseline atom interferometer experiments, detectable without altering the experimental design:  {\it massive graviton ultra-light dark matter}.

From massive gravity field theory, we identified three viable spin-2 dark matter scenarios that atom interferometers could probe. The first is the Lorentz-invariant {\bf FP} case, which includes propagating tensor, vector and scalar modes (all defined under $SO(3)$) with equal masses and propagating speeds. This case is characterized by a coupling to Standard Model fields given by Eq.~\eqref{eq:FPmasscoupling} and Eq.~\eqref{eq:couplinglightFP2}. The second is the Lorentz-violating {\bf LV1} case, where only the tensor modes propagate, with coupling characterized by the constants $\alpha^{(0)}_{\rm}$ and $\beta^{(0)}_{\rm}$ in Eqs.~\eqref{eq:couplingmatterLV} and \eqref{eq:couplinglightFP3}. The third is the Lorentz-violating {\bf LV2} case, which has a propagating tensor and scalar mode with independent masses. Its couplings appear in Eqs.~\eqref{eq:couplingmatterLV} and \eqref{eq:couplinglightFP3}, and it uniquely features a known UV completion in terms of a Higgs-like sector. 

The tensor and scalar modes are detectable by atom interferometers, coupling like gravitational waves and scalar ULDM respectively. 
The vector mode of the field (or the vector induced by derivatives of the scalar mode) would instead induce new effects, such as a delay in the propagation of the atoms leading to a phase shift. This term is suppressed in the {\bf FP} case (it includes a factor of the dark matter velocity $v_{\rm DM}\ll 1$, cf.\ Table~\ref{tab:gamma}), but may be enhanced in the ${\bf LV2}$ case.  Quite remarkably, all these effects can be parameterized in terms of a factor $\gamma_\chi$ generating a phase shift that can be measured (cf.~Eq.~\eqref{eq:signal}). 
From the sensitivity of different atom interferometer configurations to measure a given value of~$\gamma_\chi$, which we show in Fig.~\ref{fig:FP_limits}, one can easily find the constraining power for the three selected theories of massive gravity by simply using the values of $\gamma_\chi$ that characterize them. In Table~\ref{tab:gamma}, we provide such a collection of values for the couplings that we identified as more promising for the models of interest. 

In Fig.~\ref{fig:FP_limits}, we also show that the leading constraints on these signals come from fifth-force experiments and tests of the equivalence principle, which are relevant for {\it some} of the models under study. The scalar sector of the theory (even if not propagating) may cause modifications to the Newtonian potential which, in the relevant parameter space for atom interferometer experiments, are constrained from lunar laser ranging (see \cite{Dubovsky:2004sg} for how {\bf LV1} may evade these constraints).

In theories without a scalar mode in this mass range, the tensor and vector modes remain unconstrained, such that  a \SI{10}{m} scale instrument {\it{ could explore unconstrained parameter space for the first time}}.
In contrast, if the fifth-force constraints apply, the \SI{100}{m} and \SI{1}{km} baseline experiments would be needed to explore new regions of parameter space. 
The Higuchi bound may also restrict the possible parameter space for a healthy spin-2 ULDM candidate, primarily in the {\bf FP} case. However, it will likely not impact atom interferometer searches and may instead limit LISA and similar experiments. Additional constraints from superradiance warrant further study.



Networking multiple atom interferometer experiments will enhance searches. Within the spin-2 ULDM field, all terrestrial experiments are effectively co-located given the coherence lengths in the parameter space we consider. A network of multiple atom interferometers would modify the detector pattern functions $D^{i}$ and $D^{ij}$ in the elements of Table~\ref{tab:gamma}, potentially enabling the distinction of scalar, vector, and tensor signals. Any directional dependence may also be identified through daily modulations of the signal. A global network would also help mitigate the impact of local environmental noise, particularly if certain locations experience higher noise levels during certain times of the day or year~\cite{Carlton:2023ffl}.

Future work may further explore distinguishing between scalar, vector, and tensor ULDM and stochastic sources of gravitational waves. In general, we expect the spectral features of these signals to differ, necessitating extended measurement campaigns. These features have been explored for vector ULDM fields~\cite{Amaral:2024tjg} and would be complemented by a study of the theories we consider in this work. We may also extend work on super-Nyquist searches for scalar ULDM to spin-2 candidates~\cite{Badurina:2023wpk}.

Recent studies have investigated the optimization of overlap reduction functions for laser interferometer experiments in spin-2 ULDM searches~\cite{Manita:2023mnc}. Extending this analysis to atom interferometers is beyond the scope of this work, but may be an important factor in the site selection of future long-baseline atom interferometer experiments.

A future application of this work is to explore how experiments beyond atom interferometers could probe the couplings we have identified in Table~\ref{tab:gamma}. Given the established sensitivity of atomic clock comparison experiments to scalar ULDM couplings~\cite{Arvanitaki:2014faa}, we expect them to also constrain spin-2 dark matter. Additionally, other experimental setups, such as cavity experiments, may probe couplings like $\beta^{(1)}$ that are suppressed in atom interferometer experiments.

The search for dark matter is a major scientific endeavour that will shed light on open questions at the intersection of cosmology, particle physics, and gravity. 
By expanding the landscape of experimentally viable targets, our work brings spin-2 dark matter to the forefront as an exciting possibility with unique observational signatures. 
The exquisite control offered by long-baseline atom interferometers, combined with their intrinsic sensitivity to multiple dark matter couplings, makes them an extremely promising tool in this ongoing hunt for new physics.

\section*{Acknowledgements}

We are grateful to members of the AION collaboration for comments and productive discussions on this work. We are thankful to Leonardo Badurina, John Ellis, Claudia de Rham, and Federico Urban for relevant discussions and comments on the manuscript. J.C. acknowledges Lewis Croney for help in developing the ideas for this paper. 

D.B. acknowledges the support from the Departament de Recerca i Universitats from Generalitat de Catalunya to the Grup de Recerca 00649 (Codi: 2021 SGR 00649).
The research leading to these results has received funding from the Spanish Ministry of Science and Innovation (PID2020-115845GB-I00/AEI/10.13039/501100011033). This publication is part of the grant PID2023-146686NB-C31 funded by MICIU/AEI/10.13039/501100011033/ and by FEDER, UE.
IFAE is partially funded by the CERCA program of the Generalitat de Catalunya.
J.C.\ acknowledges support from a King's College London NMES Faculty Studentship. 
C.M.\ is supported by the Science and Technology Facilities Council (STFC) Grant No.\ ST/T00679X/1. 

For the purpose of open access, the authors have applied a Creative Commons Attribution (CC BY) license to any Author Accepted Manuscript version arising from this submission. 

The data supporting the findings of this study are available within the paper. 
No experimental datasets were generated by this research. 

\appendix

\section{Integration of non-propagating degrees of freedom}\label{app:dofs}

In terms of the decomposition of Eq.~\eqref{eq:splitdofs}, the LV model with the kinetic term from the FP Lagrangian \eqref{eq:FP} and massive term from Eq.~\eqref{eq:LVLag}, can be written as a 
sum of three terms~\cite{Blas:2009my}
\begin{equation}
    \mathcal{L}_\mathrm{FP} = \mathcal{L}_{t}+\mathcal{L}^0_{v}+\mathcal{L}^0_{s}, \label{eq:L0}
\end{equation}
where the Lagrangian for tensors reads
\begin{equation}
\mathcal{L}_{t}=\frac{1}{2} \left(\varphi^{\rm TT}_{i j}  \Box \varphi^{\rm TT}_{i j}-m_2^2 \varphi^{\rm TT}_{i j} \varphi^{\rm TT}_{i j}\right),
\end{equation}
while for vector one finds,
\begin{equation}
\begin{split}
    &\mathcal{L}^0_{v}=\left\{-\left(u_i-\dot{A}_i\right) \Delta\left(u_i-\dot{A}_i\right)\right.\\
&~~~~~~~~~~~~~~~\left.+\left[m_1^2 u_i u_i+m_2^2 A_j \Delta A_j\right]\right\}, \label{eq:Lv}
\end{split}
\end{equation}
and for scalars 
\begin{equation}
\begin{split}
&\mathcal{L}^0_{s}=\frac{1}{2} \left\{-6\dot\pi^2+2(2 \Psi-\pi) \Delta \pi+4\dot\pi \Delta\left(2 w-\dot\sigma\right)\right. \\
&+\left[m_0^2 \Psi^2-2 m_1^2 w \Delta w-m_2^2\left(\sigma \Delta^2 \sigma+2 \pi \Delta \sigma+3 \pi^2\right)\right. \\
&\left.\left.+m_3^2(\Delta \sigma+3 \pi)^2-2 m_4^2 \Psi(\Delta \sigma+3 \pi)\right]\right\}.
\end{split}
\end{equation}

Starting with vector degrees of freedom, the equation of motion for $u_i$ reads
\begin{equation}
    u_i=\frac{\Delta}{\Delta-m_1^2} \partial_0 A_i, \label{eq:uvsA}
\end{equation}
which is valid for all mass choices. Once integrated out in  Eq.~\eqref{eq:Lv}, one retrieves the vector component of Eq.~\eqref{eq:DOFs}.

In Eq.~\eqref{eq:A_Pi}, the vector field $A_i$ was defined in the {\bf FP} case. More generally, it is defined as
\begin{equation}
  A_i=\frac{1}{m_1} \sqrt{\frac{\Delta-m_1^2}{2 \Delta}} \widetilde{A}_i,
\end{equation}
where the $m$ terms are now $m_1$.

For the scalar modes, we distinguish three cases of interest
\begin{itemize}
    \item $m_0=0, \ m_1\neq 0$ and $m_4\neq0$. This case is relevant for the {\bf FP} case. One finds \cite{Rubakov:2004eb}
\begin{equation}\begin{split}
 &       \sigma=\frac{2}{m_4^2} \pi-\frac{3}{\Delta} \pi, ~~  w=\frac{2}{m_1^2} \dot \pi,\\ 
&    \Psi=\frac{1}{m_4^2}\left(2 \ddot \pi-2 \frac{m_2^2-m_3^2}{m_4^2} \Delta \pi+2 m_2^2 \pi\right),
\label{eq:otherdofsLI}
    \end{split}
\end{equation}
From this, the trace reads
\begin{equation}
\begin{split}
 &   \varphi^\mu_\mu=-\Psi+\Delta\sigma +3\pi =\\ &\frac{2}{m_4^2}\bigg[-\partial_0^2+\left(1+\frac{m_2^2-m_3^2}{m_4^2}\right)\Delta-m_2^2\bigg]\pi.
    \label{eq:otherdofsLITr}
\end{split}
\end{equation}
\item $m_1=0$, $m_0\neq 0$. This is relevant for  {\bf LV1}. As only tensor modes propagate, we can simply consider the effect from $\varphi^{\rm TT}_{ij}$, and do not worry about scalars or vectors. 
\item $m_0=m_1=0$. This is relevant for  {\bf LV2}. In this case, we need to also consider modifications
the kinetic part of the theory, parameterized by three constants $\lambda$, $\beta$ and $\alpha$~\cite{Blas:2014ira}. From Ref.~\cite{Blas:2014ira}, one finds that
\begin{equation}
      \alpha  \Psi=2\pi, ~ \Delta \sigma=-\pi, ~
      \Delta w\approx  \frac{1+\lambda}{\lambda + \beta}\dot \pi,
      \label{eq:otherdofsLV}
\end{equation}
where $\lambda$, $\beta$, and $\alpha$ are the parameters connected to LV in the kinetic part of the theory. The parameter $\alpha$ is particularly relevant for the scalar couplings discussed in the main text. This parameter can't vanish for well-defined theories. All the parameters parameterizing the violation of Lorentz invariance are constrained to be $\ll 1$ if the massive graviton generates ordinary gravitation~\cite{Blas:2014ira}, while they are much milder if the massive spin-2 field accounts only for dark matter \cite{Blas:2012vn,Bettoni:2017lxf}. 

\end{itemize}

\section{Directional dependence of the signal}
\label{app:directions}

 The modified transition frequencies in the elements of Table~\ref{tab:gamma} for the scalar contributions are constant. However, the polarisations of the tensor and vector modes result in a directional dependence of the signal that depends on the detector geometry. As we assume a terrestrial atom interferometry experiment that rotates with the Earth, sweeping through a randomly polarised spin-2 ULDM field, we average over all angles to find the signal. The angular dependence of the signal comes from the polarisation tensor $e_{ij}^\lambda(\theta, \phi)$ and vector $e_{i}^\lambda(\theta, \phi)$ introduced in Eq.~\eqref{eq:DMfield} and Eq.~\eqref{eq:DMfieldv} respectively. We treat the directional dependence of the signal on the dark matter velocity $\hat v_{\rm DM,\,\it i}$ in an equivalent way. 
 
 Taking the product of the detector tensor $D^{ij}$ or vector $D^{i}$ and the polarisation tensor or vector, the angular dependence is contained in detector pattern functions~\cite{Maggiore:2007ulw, Manita:2022tkl}
 \begin{equation}
     F_\lambda(\theta, \phi) =
     \begin{cases}
        D^{i}e_{i}^\lambda(\theta, \phi), & \text{ for vectors,}\\
        D^{ij}e_{ij}^\lambda(\theta, \phi), & \text{ for tensors.}
     \end{cases}
 \end{equation}
 Averaging over all polarization directions then gives the geometric factor 
\begin{equation}
    \mathcal{R}^\chi = \langle F_{\lambda,\,\chi}^2\rangle ,
\end{equation}
where $\chi=v,t$ and we assume each of the polarisations $\lambda$ has the same factor $\mathcal{R}^\chi$ when averaging over sky location $(\theta,\phi)$ and polarisation angle $\psi$. We calculate the spatial average as
\begin{equation}
        \langle ...\rangle = \int_0^{\pi} \frac{\mathrm{d}\psi}{2\pi}\int_0^\pi \frac{\mathrm{d}\theta}{2\pi}\,\sin\theta\int_0^{2\pi}\mathrm{d}\phi\, \left(...\right).
\end{equation}
\\
We expect LIGO and LISA to be only sensitive to the tensor effects we have derived.
For the LIGO experiment with two perpendicular laser arms, the geometric factor is~\cite{Maggiore:2007ulw, Robson:2018ifk}
\begin{equation}
    \mathcal{R}^t_\mathrm{LIGO} = 1/5.
\end{equation}
For LISA there is a more complex form
\begin{equation}
    \mathcal{R}^t_\mathrm{LISA} = \frac{3}{10} - \frac{507}{5040}\left(\frac{f}{f_*}\right) + ...\,
\end{equation}
where $f_* = 19.09$~mHz is the peak frequency~\cite{Robson:2018ifk}. 

Vertical, fountain-configuration atom interferometer experiments such as AION-10~\cite{Badurina:2019hst} and MAGIS-100~\cite{MAGIS-100:2021etm}, will probe one direction of the spin-2 ULDM field. When detecting tensor modes, the geometric factor is identical to that of a resonant bar detector which also probes only one direction of a field~\cite{Maggiore:2007ulw}. For detecting vector ULDM, we refer to previous work i.e.,~\cite{Caputo:2021eaa, Amaral:2024tjg}.

Summing over the contributions to the signal from each polarisation we calculate the factors 
\begin{equation}
    \begin{split}
        \sum_\lambda\mathcal{R}^v_{\rm AI} &=   \sum_\lambda\left\langle \left(D^{i} e^\lambda_{i}\right)^2\right\rangle= \frac{2}{3},\\
        \sum_\lambda\mathcal{R}^t_{\rm AI} &=  \sum_\lambda\left\langle \left(D^{ij} e^\lambda_{ij}\right)^2\right\rangle= \frac{8}{15},
    \end{split}    
\end{equation}
which we would substitute in the expressions in Table~\ref{tab:gamma} when calculating the sensitivity.

\bibliography{References.bib}

\end{document}